\documentclass[
  aip,
  rsi,
  amsmath,amssymb,
  superscriptaddress,
  reprint,
  numerical,
  twocolumn,
  longbibliography
]{revtex4-2}
\usepackage{graphicx}% Include figure files
\usepackage{dcolumn}% Align table columns on decimal point
\usepackage{bm}% bold math
\usepackage{physics}
\usepackage[normalem]{ulem}
\usepackage{siunitx}
\usepackage{hhline}
\usepackage{url}
\usepackage{caption}
\usepackage{subcaption}
\usepackage{mathtools}
\usepackage{amsmath,amssymb}
\usepackage{mhchem}
\usepackage[dvipsnames]{xcolor}
\usepackage[
  colorlinks=true,
  linkcolor=Blue,
  citecolor = Blue,
  urlcolor = Blue,
  unicode
  ]{hyperref}
\usepackage{natbib}
\usepackage[nameinlink]{cleveref}
\crefname{equation}{Eq.}{Eqs.}
\Crefname{equation}{Equation}{Equations}
\crefname{figure}{Fig.}{Figs.}
\Crefname{figure}{Figure}{Figures}

\usepackage{caption}
\usepackage{enumerate}

\usepackage[utf8]{inputenc}
\newcommand{\II}{I\hspace{-0.05em}I}

\begin{document} 

\title{Investigations of optical aberration on quantum diamond microscopy toward high spatial resolution and sensitivity
\vspace{20pt}\\}
\author{Shunsuke Nishimura}
\email{shunsuke.nishimura@phys.s.u-tokyo.ac.jp}
\affiliation{Department of Physics, The University of Tokyo, Bunkyo-ku, Tokyo, 113-0033, Japan}
\author{Moeta Tsukamoto}
\affiliation{Department of Physics, The University of Tokyo, Bunkyo-ku, Tokyo, 113-0033, Japan}
\author{Kento Sasaki}
\affiliation{Department of Physics, The University of Tokyo, Bunkyo-ku, Tokyo, 113-0033, Japan}
\author{Kensuke Kobayashi}
\email{kensuke@phys.s.u-tokyo.ac.jp}
\affiliation{Department of Physics, The University of Tokyo, Bunkyo-ku, Tokyo, 113-0033, Japan}
\affiliation{Institute for Physics of Intelligence, The University of Tokyo, Bunkyo-ku, Tokyo, 113-0033, Japan}
\affiliation{Trans-Scale Quantum Science Institute, The University of Tokyo, Bunkyo-ku, Tokyo, 113-0033, Japan}

\begin{abstract} 
  Quantum diamond microscopy (QDM), which employs nitrogen-vacancy (NV) center ensembles, is a promising approach to quantitatively imaging magnetic fields with both high resolution that approaches the diffraction limit and a wide field of view. 
  The commonly adopted setups of QDM capture the photoluminescence through transparent diamonds, which inevitably entail aberrations---optical errors that degrade the optical resolution and contrast of the obtainable image. In this study, we delve into the impact of optical aberrations, focusing on their dependence on diamond thickness. 
  We first introduce a rigorous model [B. Richards, \emph{et al.}, \href{https://doi.org/10.1098/rspa.1959.0200}{Proceedings of the Royal Society of London. Series A.
  Mathematical and Physical Sciences 253, 358--379 (1959)}, J. Braat, \emph{et al.}, \href{https://doi.org/10.1364/JOSAA.20.002281}{J. Opt. Soc. Am. A 20, 2281--2292 (2003)}.] of diffraction that incorporates aberrations, producing the NV center optical image. We confirm that this model accurately reproduces the confocal images of single NV centers obtained at various depths in diamonds. 
  Extending this model to a wide-field microscope, we find that the model also accurately  reproduces the USAF 1951 resolution test chart obtained through diamonds of various thicknesses. Based on these investigations, we quantitatively assess the consequent resolution constraints and propose thinning the diamond as a viable solution. 
  We present a robust method to quantitatively ascertain resolution in optical systems influenced by aberrations caused by ray transmission through diamonds. For instance, for a typical microscope with an objective lens of $\mathrm{NA}=0.7$, the diffraction limit is achievable through diamonds that are $30~\si{\micro m}$ thick, and a resolution of $1~\si{\micro m}$ is obtained through diamonds that are $100~\si{\micro m}$ thick. 
  Those results opens up avenues for enhanced performance in QDM.
\end{abstract}
\maketitle

\section{Introduction}\label{sec:intro}
\begin{figure}[b]
  \includegraphics[width = \hsize]{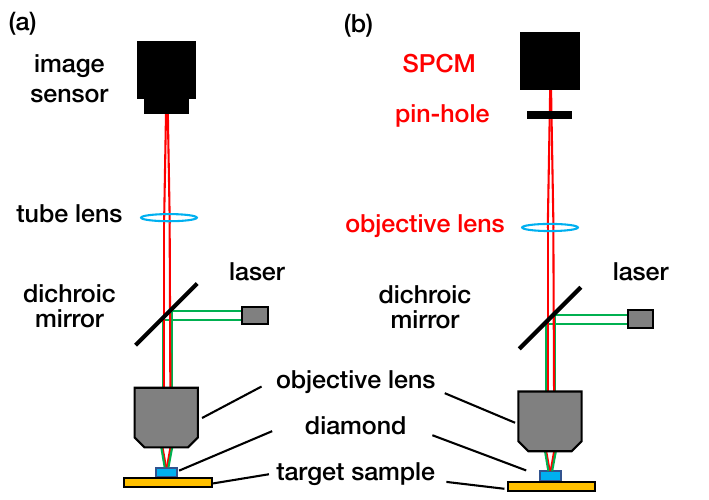}
  \caption{
      (a) Schematic of the wide-field microscope used for capturing the images in \cref{fig:imaging}.
      (b) Schematic of the confocal microscope employed in \Cref{sec:confocal}. Components that differ from those in (a) are highlighted in \textcolor{red}{red}.
  }
  \label{fig:microscopes}
\end{figure}

Quantum diamond microscopy (QDM) is an emerging approach to quantitatively imaging magnetic field using an ensemble of nitrogen vacancy (NV) centers in diamond. QDM boasts high accuracy in measuring the magnetic field and offers a combination of high spatial resolution, which approaches the diffraction limit (up to several hundreds of nm), and a wide field of view (as much as several hundreds of \si{\micro m}).
Recently, QDM has been primarily applied in the fields of condensed matter physics\cite{scholten2021widefield}, such as the current distribution in materials\cite{tetienne2017quantum,scott2019imaging,ku2020imaging,mizuno2020}, the magnetic properties\cite{scholten2022imaging,ghiasi2023}, and superconductors\cite{acosta2019color,scott2020laser,sn2023}, and also in the field of biology\cite{kucsko2013, gelnn2015single-cell}.
The spatial resolution and sensitivity to the stray field from a magnetic material of QDM are constrained by the distance between the target and the sensor, as well as the performance of the optical system \cite{scholten2021widefield}.
In many studies utilizing ensemble NV centers, the primary consideration has been to minimize the distance between the target and the sensor.
Therefore, measurements are often performed with the diamond being directly attached to the sample, and the photoluminescence (PL) from the NV center enters the aperture of the lens through the transparent diamond (see, for example, Refs. \onlinecite{tetienne2017quantum,scott2019imaging,mizuno2020,scholten2022imaging,ghiasi2023,acosta2019color,scott2020laser}), as illustrated in the configuration in \cref{fig:microscopes}(a).
However, this configuration is not optically ideal due to the existence of diamond, leading to optical aberrations \cite{born_wolf}. These aberrations, in turn, can result in a decrease in both the contrast and resolution of the PL from the NV ensemble.

\begin{figure*}[t]
  \includegraphics[width = \hsize]{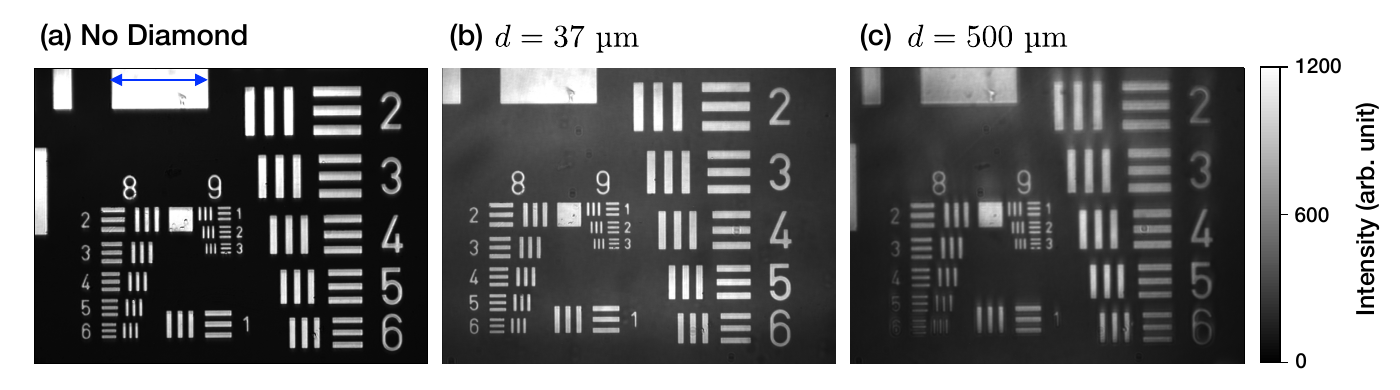}
  \caption{Images of the USAF 1951 resolution test chart\cite{usaf1951} captured using a CMOS camera.
  (a) Image captured without placing a diamond. The blue double-sided arrow indicates the scale of Group 6, Element 2 of the USAF chart,
  which has an actual size of $34.8~\si{\micro m} = 2.5 \times 2^{-(6 + 1/6)}$.
  (b) Image on which a diamond of thickness $37~\si{\micro m}$ is placed.
  (c) Image on which a diamond of thickness $500~\si{\micro m}$ is placed.
  All measurements utilize the reflection of green laser light, and in (b) and (c), the background is subtracted.
  The dimming observed at the edges of the field of view in (c) is influenced by the change in the laser spot size due to the refraction by diamond.}
  \label{fig:imaging}
\end{figure*}

We provide a brief illustration of the effect of aberration from the measurement results. 
\Cref{fig:imaging} shows the images of a USAF chart captured under different conditions. 
In \cref{fig:imaging}(a), the pattern is measured without a diamond.
In \hyperref[fig:imaging]{Figs.~\ref{fig:imaging}(b) and \ref{fig:imaging}(c)} , on the other hand, 
the pattern is measured through diamonds. 
Specifically, diamonds of $37\si{\micro\meter}$ and $500~\si{\micro\meter}$ thickness are used, respectively, placed atop the USAF chart.
Each measurement is conducted under the same optical system, with a $532~\si{nm}$
laser serving as the light source, uniformly irradiating the entire field of view, and the USAF chart pattern's reflection is measured.
Despite identical optical compensation, the results in \hyperref[fig:imaging]{Figs.~\ref{fig:imaging}(b) and \ref{fig:imaging}(c)} exhibit
a reduced signal-to-noise ratio (SNR) compared to that in \cref{fig:imaging}(a). Moreover, noticeable blurring and bleeding of the optical image are observed at the corner of view in \cref{fig:imaging}(c).

This phenomenon is attributed to the aberration caused by the focusing ray passing through a diamond \cite{hell1993aberrations,simmonds2011abberation,sun2014effects},
a material with a high refractive index ($n\sim 2.42$).
Typical commercial lenses are calibrated by assuming that the focusing ray propagates solely through media with a constant refractive index, such as air ($n = 1$) or immersion oil ($n = 1.5$). Consequently, when using such an objective lens,
aberrations arise in the focusing rays passing through media with different refractive indices.
The aberrations, depending on the relative optical distance of the passing trace, lead to blurring and bleeding in optical images.

This effect was pointed out and has been experimentally confirmed in confocal systems by \citet{hell1993aberrations}.
The same phenomenon has also been observed in QDM measurements\cite{scholten2021widefield}, 
but no comprehensive quantitative discussion has been conducted regarding the achievable resolution in real systems of QDM.
Moreover, commercial single-crystal diamond substrates typically have a thickness of $300-500~\si{\micro\meter}$. 
Given this thickness and the high refractive index of diamond, there arises a significant aberration ($\sim 10\lambda$) 
compared to the wavelength $\lambda$ of the transmitting wave. 
Such an aberration compromises the resolution \color{black}
---accompanied by artifacts potentially even beyond trivial blurring\cite{Scholten2022aberration}---
\color{black}
and sensitivity of the QDM.

Let us consider several measures for suppressing aberration. 
The aberration under consideration arises from the difference between the compensated optical path, 
assuming the medium to be air, and the optical path when transmitting through a diamond. 
The simplest method, therefore, is to use a thin diamond. Apart from this, there are three alternative measures:
\begin{enumerate}
  \item Unify the media in the optical path of converging light with a constant refractive index to eliminate the aberration.
  \item Design an objective lens that is corrected by taking into account the aberration\cite{philipp2019diffraction}.
  \item Adjust the phase for each pixel to perform the correction \cite{kner2010high-resolution} according to its value determined by modeling the wavefront aberration function or directly measuring.
\end{enumerate}
However, all three methods encounter significant difficulties when applied to transmission through a diamond, primarily due to technical reasons. 
In the case of the first method, the refractive index of diamond is high at $n = 2.42$ around $532~\si{nm}$ of visible light. 
There are no liquid media with such a high refractive index
\footnote{For instance, the upper limit for the refractive index of commercial high refractive index contact fluid is approximately $n~\sim 1.78$ \cite{shimadzu}}, 
and embedding the solid immersion lens is the only viable option \cite{huang2019monolithic}. 
However, it is physically impossible to achieve a wide field of view with this method, and it also requires advanced manufacturing techniques, lacking in experimental simplicity.
As detailed later in \Cref{sec:widefield}, in the second case, achieving the diffraction limit becomes challenging even when transmitting through very thin diamonds. 
The optical system fails to perform as anticipated if the diamond's thickness changes only slightly, leading to a lack of versatility. 
In the third case, correcting aberrations is not easy because the quantity of aberration surpasses the correctable wavelength of wavefront aberration when the diamond is thick. Consequently, to attain a high resolution up to the diffraction limit, which is the theoretical maximum for optical resolution, thinning the diamond appears to be the most practical solution.

In this study, we quantitatively investigate how resolution changes with the diamond of different thickness, 
employing both a theoretical model and measurements taken with typical microscopes. 
First, using a confocal system [\cref{fig:microscopes}(b)], we acquire the photoluminescence (PL) image of single NV centers located at different depths from the surface. 
NV centers, being atomically scaled, serve as optically ideal point light sources and the optical image of a single NV center spreads solely due to diffraction effects. 
We thus compare the experimentally obtained three-dimensional spatial distribution of PL intensity, or the point spread function (PSF), 
with the calculation from a diffraction integral model incorporating an aberration function. 
This comparison confirms that the model accurately reproduces the change in the PSF due to the aberration effects dependent on the thickness of the transmitted diamond. 

With this confirmation as our basis, we subsequently apply the diffraction model to the case of a wide-field microscope, which is more significantly influenced by aberration.
By comparing the USAF chart's optical images derived from this calculation with the measurement results depicted in \cref{fig:imaging}, 
we can verify the effectiveness of the diffraction model for a wide-field microscope. 
Based on these results, we calculate the permissible thickness of diamond required to reach the diffraction limit from the diffraction model under the condition of typical optics adopted in QDM. 
As a result, we ascertain that when a general optical system with numerical aperture $\text{(NA)} = 0.7$ is employed, the thickness of the diamond necessary to achieve the diffraction limit is approximately 40~\si{\micro m}. 
To achieve a resolution of 1~\si{\micro m} across the field of view, the thickness of the diamond within the optical path should be limited to 100~\si{\micro m} or less.
Moreover, when using a commercial diamond of 500~\si{\micro m}, the resolution is approximately 1.5~\si{\micro m} near the view center and about 3~\si{\micro m} at the edge of the field of view (70~\si{\micro m} away from the view center). 

This study presents a procedure to quantitatively determine resolution in an optical system affected by aberrations due to ray transmission through a diamond.
Our model provides guidelines for designing optics of QDM with higher resolution and sensitivity.

This paper is organized as follows.
Sections \ref{sec:diffraction} through \ref{sec:calculation} describe the diffraction model applied in this study for the evaluation of the PSF.
In \Cref{sec:diffraction}, we present a comprehensive overview of the theory. In \Cref{sec:aberration}, the aberration function is defined and the aberration function for light traversing a plane-parallel plate is introduced.
Additionally, we provide the general representation of the diffraction integral incorporating aberration.
In \Cref{sec:calculation}, we elucidate the numerical calculation method utilized in this research.
The experimental results are presented in Sections \ref{sec:confocal} and \ref{sec:widefield}.
In \Cref{sec:confocal}, we measure the optical image of a single NV center placed at different depths using a confocal microscope, and compare the changes in the PSF at varying depths from the surface of the host diamond crystal.
Subsequently, we calculate the PSF from the diffraction integral, as illustrated up to \Cref{sec:calculation}, and validate our theoretical model by comparing the calculated results with the measurements, thereby evaluating the resolution.
Additionally, we investigate the impact of aberration on the intensity of the excitation light based on the excitation rate of the NV center, as detailed in \Cref{subsec:laser_intensity}.
In \Cref{sec:widefield}, we extend the application of this theory to the case of a wide-field microscope and assess the allowable thickness corresponding to each desired resolution considering the field of view.
Finally, we present our conclusions in \Cref{sec:conclusion}.

\section{Diffraction theory}\label{sec:diffraction}
This section provides an overview of diffraction theory. Diffraction is a phenomenon where a propagating wave changes its direction and interferes around an obstacle when the wave travels around it [\citet[][Chapter VIII, Section 4]{born_wolf}]. 
Due to the effect of diffraction, it always has a spread that depends on its wavelength. 
This sets a limit to the optical resolution of an image, known as the diffraction limit. 
To ascertain the limit, we introduce a quantitative description of the diffraction effect. 
This model serves our final goal of evaluating resolution by the spatial distribution of PL, namely PSF, by considering the vectorial nature of traveling electromagnetic fields given in \Cref{subsec:rigorous_diffraction} and the effect of aberration described in \Cref{sec:aberration}.

\subsection{Diffraction integral of spherical converging ray through a circular aperture}
\label{subsec:circle_aperture}
\begin{figure*}[htbp]
  \includegraphics[width = \hsize]{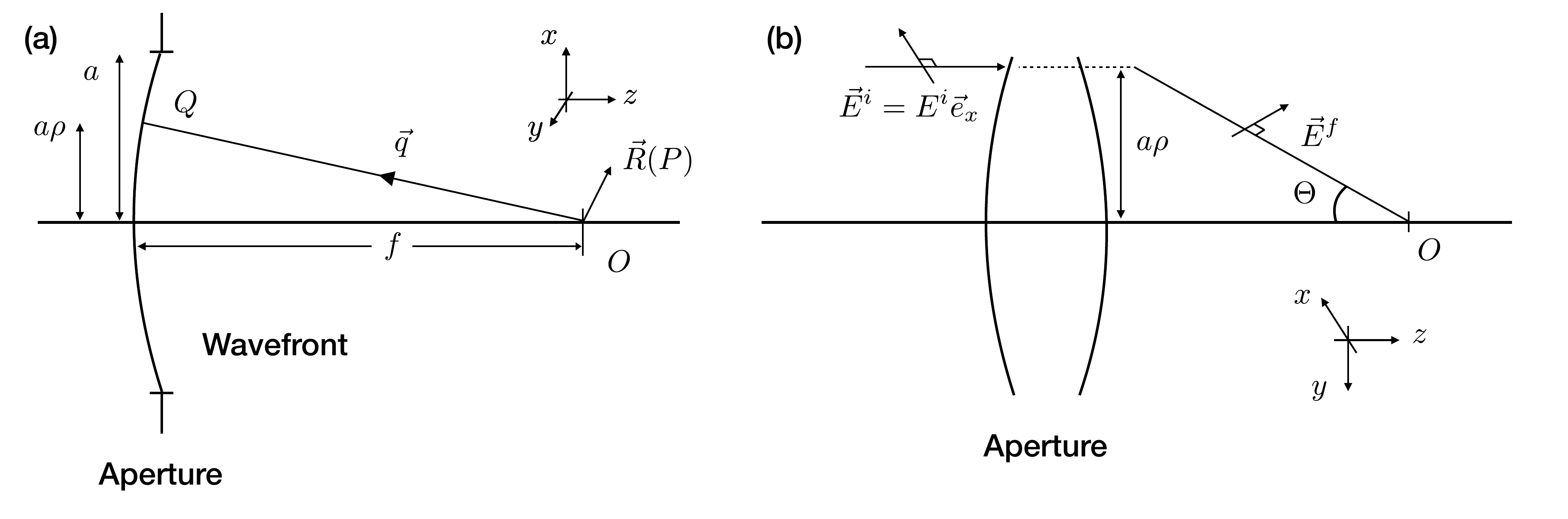}
  \caption{
    (a) Schematic representing the diffraction of focusing rays in a circular aperture.
    (b) Schematic illustrating the refraction of the incident light's electric field vector at the aperture.}
  \label{fig:diffraction}
\end{figure*}

In order to discuss the image formation in microscopes, it is necessary to consider the situation where PL from a point emitter enters into an objective lens. This situation is equivalent to the case where light rays converge after passing through a circular aperture. First, following [\citet[][Chapter VIII, Section 8]{born_wolf}], we derive the conventional diffraction integral to describe the distribution of light intensity in the vicinity of a focal point.
This derivation can be achieved by applying the scalar field theory from Kirchhoff's diffraction formula\cite{born_wolf} to a circular aperture. 
A schematic diagram representing the diffraction of converging rays in the case of a circular aperture is shown in \cref{fig:diffraction}(a). We set the origin as $O$ (i.e., the focus) and place an aperture, through which the optical axis passes the origin, at a position $f$ away from the origin.
We consider a point $P$ in the focal region at the coordinate $\vec{R}$ and the area $Q$ of the aperture, with the aperture radius denoted as $a$ and the focal length as $f$.

The amplitude of a diffracted wave resulting from a wavefront disturbance by circular aperture
can be expressed in the form of an integral over the normalized direction vector $\vec{q} \in \Omega$ from the origin $O$ to a certain point on the aperture $Q$, as follows:
\begin{equation}
U(\vec{R}) = -\frac{ifA}{\lambda} \iint_{\Omega} e^{-ik\vec{q} \cdot \vec{R}} d\Omega. \label{eq:debye}
\end{equation}
Here, $A$ is the amplitude at $Q$, $\lambda$ is the wavelength, and $k = 2\pi/\lambda$ is the wave number.
\Cref{eq:debye} is the so-called Debye integral\cite{born_wolf}.
We express each the vectors $\vec{q} = (\xi,\eta,\zeta)$ and $\vec{R} = (x,y,z)$ in cylindrical coordinates as follows,
\begin{align}
(\xi,\eta, \zeta) &= \left(\frac{a\rho\cos\theta}{f},\frac{a\rho\sin\theta}{f}, \sqrt{1-\frac{a^2}{f^2}\rho^2}\right)\notag\\
(x,y,z) &= (r\cos\psi,r\sin\psi,z)
\end{align}
Here, we use the dimensionless coordinate parameters $0 \leq\rho\leq 1,\theta,\psi \in [0, 2\pi) $. Then,
\begin{equation}
\vec{q} \cdot \vec{R} = \frac{a\rho r \cos(\theta-\psi)}{f} - z\sqrt{1-\frac{a^2}{f^2}\rho^2}.
\end{equation}
Therefore, \cref{eq:debye} is given by
\begin{widetext}
\begin{align}
  U(\vec{R}) 
  &= -\frac{is_0^2 fA}{\lambda} \int_0^1 \rho d\rho\int_0^{2\pi} d\theta
  \exp[-iks_0r\rho\cos(\theta-\psi) + ikz\sqrt{1-{s_0}^2\rho^2}]\label{eq:cir_aperture}\\
  &~\sim -\frac{is_0^2 fA}{\lambda} e^{iu/s_0^2}\int_0^1 \rho d\rho\int_0^{2\pi} d\theta
  \exp[-iv\rho\cos(\theta-\psi) - \frac{1}{2}iu \rho^2].\label{eq:cir_aperture_approx}
\end{align}
\end{widetext}
where $s_0,u$ and $v$ are the dimensionless parameters defined as follows,
\begin{equation}
  s_0 = \frac{a}{f},\quad u = k {s_0}^2 z,\quad 
  v = k s_0 r,
\end{equation}
and the integral variable $d\Omega$ is rewritten with the coordinate variables as
\begin{equation}
  d\Omega = \frac{a^2\rho d\rho d\theta}{f^2}.
  \end{equation}
Furthermore, for small $s_0\rho$, the following approximation, can be introduced up to the second order,
\begin{equation}
\sqrt{1-s_0^2 \rho^2}~\sim 1 - \frac{1}{2} s_0^2 \rho^2. \label{eq:radiometric}
\end{equation}
By using this approximation, \cref{eq:cir_aperture} is often expressed in a more computationally convenient form, \cref{eq:cir_aperture_approx}.

Although this method relies on approximations, according to [\citet[][Chapter VII, Section 4]{born_wolf}], this scalar theory can apply to general scalar waves and also to traveling electromagnetic waves in many cases. Specifically, it is apt in cases where unpolarized quasi-monochromatic light is passing through an aperture that is not excessively large.
In these cases, the intensity, defined as
\begin{equation}
I = \frac{c}{4\pi} |\langle \vec{E} \land \vec{H} \rangle|, 
\end{equation}
where $\land$ means the cross product, represents the time-averaged energy flux density. 
This can also be expressed as $I= |\phi(\vec{R})|^2$ using the complex scalar wave function $\phi(\vec{R})$ of a single harmonic oscillator.
A detailed proof of this assertion can be found in \citet{wolf1959scalar,theimer1952foundation}.
Therefore, for understanding the intensity distribution in most optical systems, it is sufficient to calculate \cref{eq:cir_aperture_approx}.

However, the assumption that the aperture is not very large is not guaranteed in systems with a high NA \cite{wolf1959electromagnetic, wolf1959scalar}.
Specifically, the vectorial nature of diffracted light cannot be neglected when the NA exceeds 0.6\cite{braat1997analytical, braat2003extended, van2010extended}.
Additionally, it has also been pointed out that the choice of wavefront, specifically the parabolic approximation of \cref{eq:radiometric}, does not hold when the NA is large \cite{van2010extended}.
Furthermore, deviations from these approximations become particularly substantial when the aberrations, which are described later in \Cref{sec:aberration}, are significant \cite{braat2003extended, van2010extended}.
Therefore, it is crucial to account for and correct these issues.
\footnote{In fact, for example, \citet{hell1993aberrations} constructs a model that takes into account the effects of the vectorial nature of the electric field, even though it uses a paraxial approximation, and successfully explains experimental results.} 
Having discussed the scalar theory of diffraction, we now proceed to extend this to a vectorial expression in the next subsection. 
This is crucial for applications in systems with large apertures, where the scalar theory's approximations may not hold.

\subsection{Vectorial diffraction integral of converging spherical waves}\label{subsec:rigorous_diffraction}
We introduce the refined expression from \cref{eq:cir_aperture}, obtained in \Cref{subsec:circle_aperture}, 
to provide a more rigorous formulation for the diffraction integral of an electromagnetic wave. 
Specifically, we apply the scalar theory explained in \Cref{subsec:circle_aperture} to each component of the refracted electromagnetic field vector, 
thereby deriving the diffraction expression for each electric field component. 
Furthermore, we present it in a form that does not rely on the approximation of \cref{eq:radiometric}. 
The derivation follows [\citet{wolf1959electromagnetic}, \citet{braat2003extended}].

\cref{fig:diffraction}(b) provides a schematic that illustrates how each component of the electric field vector varies in the direction of propagation due to refraction. 
We focus on the component of the focusing rays that head toward the focus at an angle $\Theta$. 
Suppose the light passing through the aperture covers an area of the annulus $d S^i$ and the electric field amplitude upon incidence is $E^i$. 
We denote the annulus after passing through the aperture as $d S^f$ and the amplitude of the electric field there as $E^f$. 
From the intensity law of geometrical optics, we have
\begin{equation}
(E^i)^2dS^i = (E^f)^2 dS^f.
\end{equation}
Also, since $dS^i$ corresponds to the axial projection of $dS^f$ at angle $\Theta$, we have
\begin{equation}
dS^i = \cos\Theta dS^f.
\end{equation}
Therefore, when considering the diffraction of linearly polarized light, the amplitude after passing through the aperture becomes
\begin{equation}
E^f = E^i \sqrt{\cos\Theta}. \label{eq:geo_optic_int}
\end{equation}
Subsequently, we determine the amplitude and direction of the electric field vector $\vec{E}^f$ after passing through the aperture $d S^f$. With the assumption that the propagation of each component is independent, we can consider that the incident light is polarized in the {$x$-axis} direction, 
\begin{equation}
\vec{E}^i = E^i \vec{e}_x, \label{eq:Ei}
\end{equation}
without loss of generality. 
For a given coordinate of the aperture $(\rho,\theta)$, the angle $\Theta$ is given by $\Theta = \arcsin(s_0\rho)$. 
We also denote the cylindrical coordinate bases in the image space as
\begin{gather}
  \vec{e}_\rho =  \cos\theta \vec{e}_x  + \sin\theta \vec{e}_y, \notag\\ 
  \vec{e}_\theta = \vec{e}_z\wedge \vec{e}_\rho = -\sin\theta \vec{e}_x  + \cos\theta \vec{e}_y.
\end{gather}
Assuming that the electric field vector is always orthogonal to the direction of light, the direction of $\vec{E}^f$ 
is rotated by $R_{\vec{e}_\theta}(-\Theta)$ compared to the direction of $\vec{E}^i$,  where $R_{\vec{n}}(\phi)$ symbolizes the operator representing the rotation by $\phi$ around the axis $\vec{n}$.
Using \cref{eq:geo_optic_int,eq:Ei}, we obtain
\begin{align}
  \vec{E}^f 
  &= \sqrt{\cos\Theta} R_{\vec{e}_\theta}(-\Theta)\vec{E}^i \notag\\
  &= E^i \sqrt{\cos\Theta}
  [(\vec{e}_x\cdot \vec{e}_\rho)\vec{e}_{\rho'} +
  (\vec{e}_x\cdot \vec{e}_\theta) \vec{q} \wedge \vec{e}_{\rho'}]\,
\end{align}
where 
\begin{gather}
  \begin{aligned}
  \vec{e}_{\rho'} &= R_{\vec{e}_\theta}(-\Theta)\vec{e}_{\rho}\\
  &= \cos\Theta  (\cos\theta \vec{e}_x  +\sin\theta \vec{e}_y) +\sin\Theta \vec{e}_z,
  \end{aligned}
  \notag\\
  \vec{q} \wedge \vec{e}_{\rho'} = \vec{e}_{\theta}.
\end{gather}
The amplitude and direction of the electric field in the focal region represented by $\vec{E}$ is obtained by the Debye integral independently for each component of $\vec{E^f}$, similarly as performed in the scalar theory [\Cref{subsec:circle_aperture}], i.e., calculating
\begin{equation}
  \vec{E} = -\frac{if}{\lambda}\iint_\Omega e^{-ik\vec{q}\cdot\vec{R}}\vec{E}^{f}(\rho,\theta)d\Omega.
\end{equation}
Consequently, we obtain the expression for the electric field vector in the proximity of the image point as \cref{eq:Ez},
\begin{widetext}
\small
\begin{align}
  E_x & = -A\int_0^1 d\rho \int_0^{2\pi} d\theta s_0^2 \rho \frac{\sqrt{1 - s_0^2 \rho^2} + (1-\sqrt{1 - s_0^2 \rho^2})\sin^2\theta}{\sqrt[4]{1 - s_0^2 \rho^2}} \exp(-iv\rho\cos(\theta-\psi) + iu{\sqrt{1 - s_0^2 \rho^2}}) \notag\\%\label{eq:Ex}\\
  E_y & = A\int_0^1 d\rho \int_0^{2\pi} d\theta s_0^2 \rho \frac{(1-\sqrt{1 - s_0^2 \rho^2})\cos\theta\sin\theta}{\sqrt[4]{1 - s_0^2 \rho^2}} \exp(-iv\rho\cos(\theta-\psi) + iu{\sqrt{1 - s_0^2 \rho^2}})  \notag\\%\label{eq:Ey}\\
  E_z & = A\int_0^1 d\rho \int_0^{2\pi} d\theta s_0^3 \rho^2 \frac{\cos\theta}{\sqrt[4]{1 - s_0^2 \rho^2}} \exp(-iv\rho\cos(\theta-\psi) + iu{\sqrt{1 - s_0^2 \rho^2}}). \label{eq:Ez}
\end{align}
\normalsize
\end{widetext}
Here, $A$ is redefined to
\begin{equation}
  A = \frac{iE^i s_0^2f}{\lambda}.
\end{equation}
Note that this expression corresponds to the replacement of the following terms that appear in [\citet[][Eq. (2.26)]{wolf1959electromagnetic}]:
\begin{align*}
\sin\theta \to s_0\rho,&\quad \phi\to\theta \\
\sin\theta (\cos\theta)^{1/2}d\theta &\to \frac{s_0^2 \rho }{\sqrt[4]{1-(s_0\rho)^2}}d\rho\\
\bm{k}\cdot\bm{x} \to  k r \rho \cos(\theta - \psi) &+ kz\sqrt{1 - (s_0\rho)^2}
\end{align*}
Furthermore, in \citet{braat2003extended}, 
the authors performed an analytical integration with respect to $\theta$, expressing it as a series of Bessel functions.
While such an analytical series representation is rigorous and widely recognized, 
in this paper we retain the integral of $\theta$ and implement numerical integration for technical reasons. 
\footnote{In this study, we use an adaptive step numerical integration method, which allows us to algorithmically determine the step width according to the tolerance of calculation accuracy by expressing it as a multiple integral.}

\begin{figure*}[tbp]
  \includegraphics[width = \hsize]{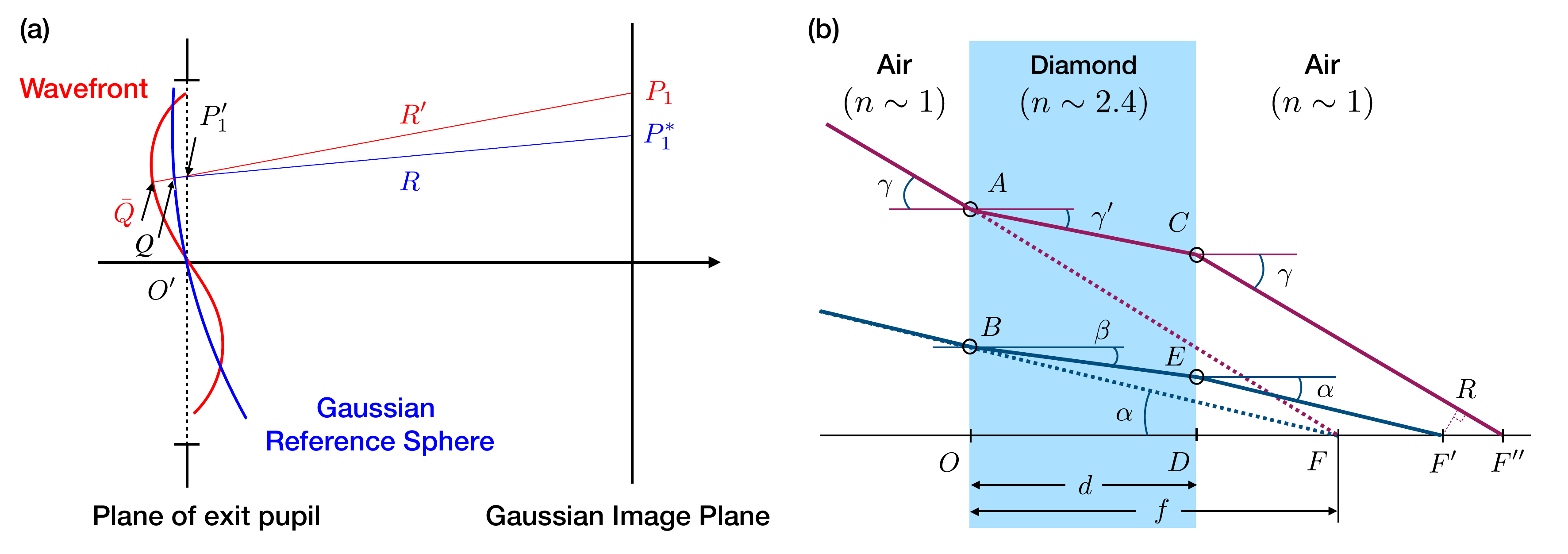}
  \caption{(a) Schematic representation of wavefront aberration. The aberrated wavefront (red line) deviates from the ideal spherical surface [the Gaussian reference sphere (blue line)], causing a phase error.
  (b) Optical path for a tilted incidence of the focusing ray through the diamond. 
  $R$ corresponds to the projection of point $F'$ onto the red solid line (incident light path with angle $\gamma$). 
  The center of the aperture lies on the line $FB$.}
  \label{fig:wavefront}
  \end{figure*}

Finally, for random polarization and propagation by electromagnetic modes, the intensity of light is represented as
\begin{equation}
I \propto | E_x |^2 + | E_y |^2 + | E_z |^2.
\label{eq:I_E}
\end{equation}
The expressions of the diffraction integral [\cref{eq:Ez,eq:I_E}] 
hold when the light passing through the optical system is represented by ideal spherical waves.

In a practical optical system, the ideal spherical wave is hard to achieve, and there are often cases where the phase-front surface deviates from the spherical wave due to various factors. 
Such a phenomenon is called an \textit{aberration}, 
and its influence must be incorporated into the diffraction integral. 
In next \Cref{sec:aberration}, we give the definition of aberration and the aberration function that characterizes the aberration 
when the light transmits through a diamond. 
The diffraction integral in the form that includes this aberration function is given,  and the theoretical model used in this paper follows that expression.

\section{Aberration}\label{sec:aberration}

Aberration refers to the phenomenon where a wavefront deviates from its ideal propagating mode due to various errors in the optical system. 
When aberrations are significant, it is possible to perform approximate calculations based on geometrical optics. 
The foundation for these calculations lies in Seidel's aberration theory\cite{seidel}, which summarizes the third-order aberration terms. 
Although Seidel's aberration theory is purely based on geometrical optics, 
Nijboer and Zernike successfully explained the diffracted imaging in the presence of aberrations using a combination of 
diffraction theory and Seidel's aberration theory. \cite{born_wolf}
In this section, we mainly explain the intensity distribution around the focus using the diffraction theory based on the Nijboer--Zernike theory, 
and we use the expression of the diffraction integral rewritten according to \Cref{subsec:rigorous_diffraction}. 
This corresponds to the extended Nijboer--Zernike theory \cite{braat2003extended, van2010extended}.

\subsection{Definition of wavefront aberration}\label{sec:definition_of_wavefront}

In the diffraction theory without aberration, the wavefront of the propagating wave is considered to coincide with the phase-front surface of the spherical wave. 
Therefore, even for converging light rays directed toward a certain point, 
the phase-front surface of the propagating wave maintains an equal distance to that point. 
On the other hand, when the optical system is not ideal, the phase-front surface becomes non-spherical. 
The schematic diagram of the situation where an aberration occurs is shown in \cref{fig:wavefront} (a). 
If there is no aberration, the original wavefront is a sphere with a radius $R$ which corresponds 
to the distance between the image point $P_1^*$ and the center of the aperture $O'$ in \cref{fig:wavefront}(a).
This sphere is called the Gaussian reference sphere \cite{born_wolf}. 
The optical distance $Q\bar{Q}$ between this Gaussian reference sphere and the actual wavefront corresponds to the actual phase shift. 
The function $\Phi(\rho, \theta)$, which represents this shift depending on the aperture coordinates $(\rho, \theta)$, is called the aberration function
\cite{wolf1951diffraction,born_wolf,braat1997analytical}.

\subsection{Aberration function in ray transmitting through a plane-parallel plate}\label{subsec:planeparallel}
Based on the definition in \Cref{sec:definition_of_wavefront}, we
define the aberration function to quantify the degree of aberration
and calculate for cases when passing through a diamond.
From this point onward, we treat the diamond as a transparent plane-parallel plate with a refractive index of $n\sim 2.42$.
Following the definition in \Cref{sec:definition_of_wavefront}, 
the aberration function when light traverses a plane-parallel plate equates to the augmented optical distance caused by placing the plate in its original path. 
\Cref{fig:wavefront}(b) replicates [\citet[][Fig. 1]{braat1997analytical}] to introduce the aberration function. The blue solid line
represents the actual optical path of the principal ray [a ray that passes through the aperture center $ (\rho= 0)$] with an incidence angle $\alpha$. The aberration arises because the phase shift experienced by light transmitting
through diamond varies for each ray path with an incident angle $\gamma$ (red solid line). The aberration function
is determined by the difference in phase shift between that of a given ray and that of the principal ray.
When we set $\Phi = 0$ to the aberration function associated with the principal ray's path (blue line), the aberration function for the path represented by the red line in \cref{fig:wavefront}(b) can be derived as [\citet[][Eq.~(5)]{braat1997analytical}], 
\begin{widetext}
\begin{align}
  \Phi(\gamma,\alpha) 
  &\stackrel{\text{def}}{=} ( n{{AC}} + {{CR}} -{{AF}} )- ( n{{BE}} + {{EF'}} -{{BF}} ) \tag{\citet[][Eq.~(4)]{braat1997analytical} }\\
  &= \qty( n\frac{d}{\cos\gamma'} + \frac{{{DF''}}}{\cos\gamma} - {{F'F''}}\cos\gamma  -\frac{f}{\cos\gamma} )- \qty( n\frac{d}{\cos\beta} + \qty(f-\frac{\tan\beta}{\tan\alpha}d)\frac{1}{\cos\alpha} -\frac{f}{\cos\alpha}) \notag\\
  &= \qty[ n\frac{d}{\cos\gamma'} + \qty(f-\frac{\tan\gamma'}{\tan\gamma}d)\frac{1}{\cos\gamma} - \qty(\frac{\tan\beta}{\tan\alpha} - \frac{\tan\gamma'}{\tan\gamma} ) d\cos\gamma  -\frac{f}{\cos\gamma} ]
  - \qty( \frac{nd}{\cos\beta}- \frac{d}{n\cos\beta}) \notag\\
  &= \frac{d}{n}\qty(\frac{1}{\cos\beta} - \frac{\cos\alpha}{\cos\beta}\cos\gamma) -nd\qty(\frac{1}{\cos\beta} - \cos\gamma').\tag{\citet[][Eq.~(5)]{braat1997analytical} }\\
  \label{eq:opt_path}
\end{align}
\end{widetext}
Here, from Snell's law, we have
$\sin\gamma = n\sin \gamma',\,\sin\alpha = n\sin\beta$, and we denote ${OD} = d, \, {OF} = f$. 
Since the overall phase shift does not affect the intensity, 
the phase shift of the ray passing through the center of the aperture has been subtracted from the total.
From \cref{eq:opt_path}, the aberration function $\Phi$ depends on the incident angle $\gamma$ and $\alpha$, and is proportional to $d$.
Furthermore, $\Phi$ does not depend on $f$ nor the position where the plate is put behind the aperture.
\color{black}
In other words, the aberration function is not affected by a finite gap between the plate surface and the measured point; it depends purely on the distance that the ``converging ray'' travels.
\cite{braat1997analytical}.
\color{black}

\subsection{Specific expressions for the case of quantum microscopes using an NV ensemble}

\begin{figure*}[tbp]
  \includegraphics[width = 0.9\hsize]{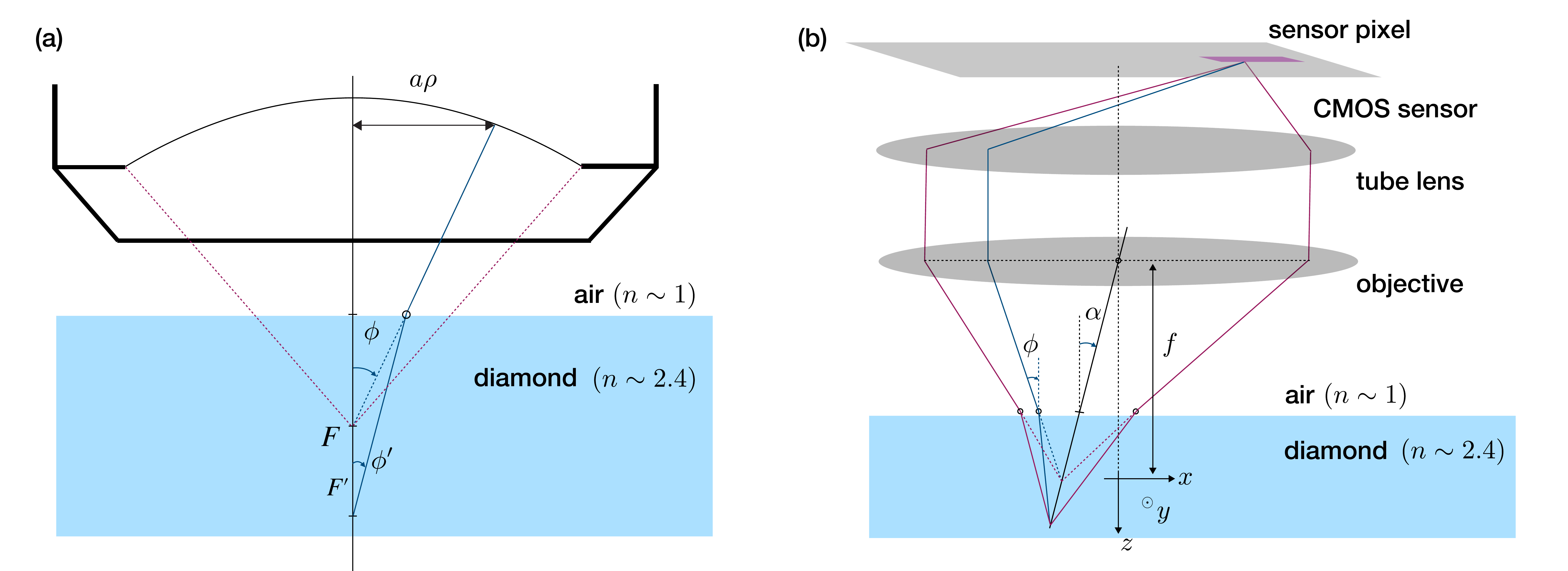}
  \caption{(a) Schematics of aberration for normal incidence, which corresponds to confocal systems.
  (b) Schematics of aberration for tilted incidence, which corresponds to wide-field microscopes with an image sensor.}
  \label{fig:normal_incidence}
\end{figure*}

The aberration function above is expressed in a formula that depends on the incident angle $\gamma$. 
Since $\gamma$ depends on the aperture coordinates $(\rho, \theta)$, we need to rewrite it in a form that explicitly reflects this dependence for the diffraction integral.
We replace the incident angle of the aberration function in each measurement 
using a confocal microscope and a wide-field microscope with an image sensor, as shown in 
\hyperref[fig:normal_incidence]{Figs.~\ref{fig:normal_incidence}}(a) and (b), respectively.

\subsubsection*{Aberration function for confocal microscopes (normal incidence)}
When observing the NV center in a diamond using a confocal microscope, in most cases, PL is measured perpendicular to the surface of the diamond.  
\Cref{fig:normal_incidence}(a) illustrates the optical path in the case of normal incidence, where the principal ray is perpendicular to the plane-parallel plate [$\alpha = \beta = 0$ in \cref{fig:wavefront}(b)]. 
This situation corresponds to the use of a confocal system [\cref{fig:microscopes}(b)] that utilizes a motorized stage, as it can be assumed that the optical axis is always aligned perpendicularly to the diamond surface when scanning the entire view area. 
For a ray specified by the aperture coordinate $(\rho,\theta)$, the angle of incidence is given by $\arcsin(s_0 \rho)$.
In this case, where the ray passes through a plane-parallel plate with refractive index $n$ and thickness $d$, the aberration function $\Phi_0$ is expressed as
\begin{align}
    \Phi_0(\rho,\theta)  &= \frac{d}{n}~(1-\cos\phi) - nd \qty(1 - \sqrt{1-\frac{\sin^2\phi}{n^2}}), \label{eq:Phi0} \\
    \phi &= \arcsin(s_0\rho),
\end{align}
obtained by setting $\alpha = \beta = 0$, and substituting $\gamma$ with $\phi$ in \cref{eq:opt_path}.
\color{black}
It is important to note that, although the aberration function does not explicitly contain $s_0$, it actually depends on the numerical aperture (NA) through the relation: $\phi = \arcsin(s_0 \rho), \, s_0 = \text{NA}.$  
Thus, if the NA is increased, the aberration becomes significant, and conversely, if the NA is decreased, the aberration is suppressed.
\color{black}

\subsubsection*{Aberration function for wide-field microscopes with an image sensor (tilted incidence)}

For wide-field microscopes with an image sensor, on the other hand, as shown in \cref{fig:normal_incidence}(b), 
the direction of incidence is tilted relative to the direction normal to the plane-parallel plate for pixels outside the view center.
This effect appears in the aberration function. 
The incident angle $\alpha$ of the principal ray is expressed depending on $(x,y,z)$ in the focal region and the focal length $f$ of the objective lens as follows,
\begin{equation}
    \alpha \sim \arctan(\frac{\sqrt{x^2 + y^2}}{f}),
\end{equation}
Then, to obtain the aberration function $\Phi$, substitute the following terms:
\begin{align*}
  \sin\phi =& s_0 \rho\\
   \sin^2\gamma\to&\sin^2\alpha (1 -\sin^2 \phi \cos^2\theta) + \cos^2\alpha\sin ^2\phi \\
    &-\sin 2\alpha\sin \phi \cos\phi \cos \theta .\tag{\citet[][Eq.~(12)]{braat1997analytical} }
\end{align*}
We obtain the aberration function $\Phi$ explicitly depending on $\rho,\theta$ and $\alpha$ as 
\begin{align}
    \Phi(\rho,\theta) =& -\frac{d\cos\alpha}{\sqrt{n^2 - \sin^2 \alpha}}(\sin\alpha\sin\phi\cos\theta + \cos\alpha\cos\phi) \notag\\
    &+ \frac{d(1-n^2)}{\sqrt{n^2-\sin^2\alpha}} + d\Biggl(n^2 - \sin^2\alpha - \cos^2\alpha\sin^2\phi \notag\\
    &- \sin^2\alpha\sin^2\phi\cos^2\theta -\sin 2\alpha \sin\phi\cos\phi\cos\theta\Biggr)^{1/2}.\notag\\
    \label{eq:Phi}
\end{align}
\color{black}
Here again, note that the aberration function $\Phi$ implicitly depends on $s_0$ (NA).
\color{black}
\subsection{Diffraction integral considering the effects of aberration and refractive index}
We provide a complete expression for the diffraction integral given in \Cref{subsec:rigorous_diffraction}, 
taking into account the effects of aberrations and the refractive index in the diamond emission. 
To incorporate the effect of aberrations in \cref{eq:Ez}, we replace each component of the electric field vector 
based on the Nijboer--Zernike theory. Specifically, we translate the phase term by the aberration function $\Phi$ as follows,
\begin{align*}
&\exp(-iv\rho\cos(\theta-\psi) + iu{\sqrt{1 - s_0^2 \rho^2}}) \notag\\
\to &\exp(ik\Phi -iv\rho\cos(\theta-\psi) + iu{\sqrt{1 - s_0^2 \rho^2}}).
\end{align*}
Moreover, within the diamond, the Gaussian reference sphere at the image point coordinates $(x, y, z)$, 
expressed as a relative distance from the origin $O$, is affected by a higher refractive index $n$ compared to vacuum. 
Specifically, following the approach of [\citet[Eq.~(2.40)]{van2010extended}], the term $u$ should be replaced as
\begin{equation}
  u{\sqrt{1 - s_0^2 \rho^2}} \to kz\frac{1-\sqrt{1-s_0^2}}{1-\sqrt{1-s_1^2}}\sqrt{1-s_1^2\rho^2}, 
\end{equation}
where $s_1 = s_0/n$. 
These substitutions give the complete expression as \cref{eq:Ez_f},
using
\begin{gather}
  u = \frac{2\pi z}{\lambda} \frac{1-\sqrt{1-s_0^2}}{1-\sqrt{1-s_1^2}},\notag\\
  v = \frac{2\pi s_0}{\lambda} \sqrt{x^2 + y^2} \notag\\
  A = \frac{iE^i s_0^2f}{\lambda},
  \label{eq:uv}
\end{gather}
\begin{widetext}
\begin{align}
  E_x & = -A\int_0^1 d\rho \int_0^{2\pi} d\theta s_0^2 \rho \frac{\sqrt{1 - s_0^2 \rho^2} + (1-\sqrt{1 - s_0^2 \rho^2})\sin^2\theta}{\sqrt[4]{1 - s_0^2 \rho^2}} \exp(\frac{2\pi i}{\lambda}\Phi -iv\rho\cos(\theta-\psi) + iu{\sqrt{1 - s_1^2 \rho^2}}) \notag\\%\label{eq:Ex}\\
  E_y & = A\int_0^1 d\rho \int_0^{2\pi} d\theta s_0^2 \rho \frac{(1-\sqrt{1 - s_0^2 \rho^2})\cos\theta\sin\theta}{\sqrt[4]{1 - s_0^2 \rho^2}} \exp(\frac{2\pi i}{\lambda}\Phi-iv\rho\cos(\theta-\psi) + iu{\sqrt{1 - s_1^2 \rho^2}})  \notag\\%\label{eq:Ey}\\
  E_z & = A\int_0^1 d\rho \int_0^{2\pi} d\theta s_0^3 \rho^2 \frac{\cos\theta}{\sqrt[4]{1 - s_0^2 \rho^2}} \exp(\frac{2\pi i}{\lambda}\Phi-iv\rho\cos(\theta-\psi) + iu{\sqrt{1 - s_1^2 \rho^2}}). \label{eq:Ez_f}
\end{align}
\end{widetext}
We denote the intensity distribution as $I$ and obtain it from numerical calculations using the aforementioned expressions \cref{eq:Ez_f,eq:uv,eq:I_E}.
Among the expressions for the diffraction integral [\cref{eq:Ez_f,eq:uv,eq:I_E}], 
the parameters that change in the experiment are the wavelength $\lambda$, the thickness $d$ of the diamond (and hence the aberration function $\Phi$), and the coordinates of the target $(x,y,z)$.
Note that $\lambda$ always represents the value in the air (vacuum). 
Hereafter, only the variables of interest are displayed as an argument of $I$ for simplicity.

\color{black}
In addition to these, with regard to the use of objective lenses with immersion liquid, special attention must be paid to numerics. 
If an oil-immersion lens is used, the PSF cannot be calculated by directly substituting $s_0 = \mathrm{NA} \sim 1.4$, 
as this results in the value inside the square root becoming negative. 
Instead, the following substitutions must be made in the diffraction integral:
\begin{equation}
  s_0 \to s = \frac{\mathrm{NA}}{n}, \, k = \frac{2\pi}{\lambda} \to k' = n\frac{2\pi}{\lambda},
\end{equation}
where $n$ is the refractive index of the immersion oil. 
Additionally, the definition $n$ in the aberration functions becomes $n_r$, which is the relative refractive index divided by the refraction index of the oil.
\color{black}
\subsection{Tolerance to aberrations}\label{subsec:tolerance}
We define the index to estimate the tolerance of optical systems to aberrations.
Specifically, we employ the Strehl ratio $S$, a widely used index to evaluate errors in optical theory \cite{born_wolf, van2010extended}. 
The Strehl ratio is defined as the ratio of the light intensity at the focus 
in the case of no aberration to the case with aberration, as follows,
\begin{equation}
S = I_f(\Phi)/I_f(\Phi = 0),
\end{equation}
where $I_f(\Phi)$ represents the intensity at the focal point (or the peak intensity) expressed in a form that explicitly depends on the aberration function $\Phi$. 
It is difficult in a general situation with aberration to precisely define the changes in resolution, such as the full width at half maximum of the actual point spread function (PSF).
If the aberration is around minimal, however, 
the PSF has approximately a $\mathrm{sinc}$ shape. If $ S > 0.8$ is satisfied, the decline in optical resolution can be suppressed to less than 4\%, 
which is practically negligible. 
Therefore, in practical optical systems, achieving the diffraction limit is often defined as satisfying $ S > 0.8$ \cite{born_wolf, van2010extended}.

\section{Numerics for the optical image of the NV center}\label{sec:calculation}
We describe the specific procedures to compute the PL intensity distribution of NV centers. 
Initially, we describe a method for calculating the PL distribution of a single NV center measured with a confocal microscope. 
Subsequently, we present calculations for the ensemble of NV centers, where the continuous optical image is influenced by diffraction effects. 
Given the recent dramatic advancements in computational capabilities, it is now feasible to perform numerical integration within a practical time frame, even for complex integrands. 
Consequently, we carry out numerical integration using the most rigorous expression possible to minimize assumptions.

\subsection{Weighting by the spectrum of NV centers}
The PL of the NV center spans a continuous range from $\lambda\sim650$~nm to $\lambda\sim750$~nm, represented by the spectrum $\sigma(\lambda)$
Therefore, the PSF is represented by the spectrum-weighted integral, as
\begin{equation}
    I_\text{NV} = \int d\lambda~\sigma(\lambda) I(\lambda).
    \label{eq:I_tot}
\end{equation}
In this study, we use a confocal microscope as described in \Cref{sec:confocal} 
to pre-measure the PL from a single NV center at a sufficiently shallow depth where the PL is free from aberration. 
We then carry out spectroscopy to calibrate the spectrum $\sigma(\lambda)$.
In subsequent numerical calculations, we use the spectrum-weighted integral given by \cref{eq:I_tot} to compute the PL intensity of NV centers.

\subsection{Calculating the PL intensity distribution of a single NV center in a confocal system}
\label{subsec:calc_confocal}
The PSF observed in a confocal system is calculated using the diffraction integral dependent on the aberration function (\cref{eq:Phi0}) caused in the case of normal incidence. 
In this context, there are additional factors that must be considered to obtain the PL intensity distribution of the confocal microscope as follows:
Firstly, the observed PSF of a single NV center must account for 
\color{black}
the diffraction on both the target NV centers' side and the
PSF of the objective lens on the Single Photon Counting Module (SPCM) side.
Here, the pinhole is modeled as a delta-function spatial filter. 
The object intensity distribution projected onto the detector plane, $O_\text{eff}(\vec{r})$, can be written using the detector-side PSF, $I_\text{detect}(\vec{r})$, if the pinhole shape (i.e., a spherical delta function $\delta(\vec{r})$) is given, as:
\begin{equation}
  O_\text{eff}(\vec{r}) = \delta(\vec{r}) I_\text{detect}(\vec{r}).
\end{equation}
The resulting confocal PSF, $I_\text{confocal}(\vec{r})$, can then be expressed using the PSFs of both sides, $I_\text{target}(\vec{r})$ and $I_\text{detect}(\vec{r})$, as \citep{Wilson1985, Pawley2006}:
\begin{align*}
  I_\text{confocal}(\vec{r}) 
  &= I_\text{detect}(\vec{r}) * O_\text{eff}(\vec{r}) \\
  &= \int d\vec{r}' I_\text{detect}(\vec{r}) \delta(\vec{r} - \vec{r}') I_\text{target}(\vec{r} - \vec{r}') \\
  &= I_\text{detect}(\vec{r}) I_\text{target}(\vec{r}).
\end{align*}
Based on this relation, we calculate the confocal image.
\color{black}

Secondly, the PL emission of the NV center itself is also dependent on the intensity of the excitation light, typically being linear.
We use a Gaussian beam\cite{orazio_svelto} as the excitation light whose intensity distribution is represented as
\begin{align}
  I_{b}(x,y,z) &= \frac{w_0 ^2}{w_z^2} \exp(-\frac{2(x^2 + y^2)}{w_z^2}),\label{eq:beam}\\
  w_z &= w_0 \sqrt{1 + \qty(\frac{z}{z_R})^2},\, z_R = \frac{nw_0}{s_0}, \notag
\end{align}
for any given point $(x,y,z)$ near the focus. Here we set the origin $O$ at the focus, and $w_z$ represents the $1/e^2$ radius of the electric field at the given $z$-plane. Furthermore, $w_0 = w_{z=0}$ denotes the beam waist radius, and $z_R$ is the Rayleigh range. We consider that NV centers PL  emission rate is linearly dependent on $I_b$.

The Gaussian beam is also subject to aberration when passing through the diamond. In the current context, the defocus term, arising from paraxial focal shift, is considered the dominant factor, as it is the lowest order aberration \cite{born_wolf}.
Since diamonds have a very high refractive index($n\sim 2.4$), paraxial focal shifts occur due to refraction, causing the {$z$-coordinate} to appear shifted upward by $(n-1)d/n$ from the actual position of the NV center. 
For wavelengths near the spectrum of the NV center, around $700$~nm, a value of $n = 2.40$ is typical\cite{turri2017index}, whereas for the wavelength of the irradiating Gaussian beam, being $532$~nm, a value of $n = 2.42$ is typical. Such differences in refractive index, causing a mismatch between the focal points of irradiation and luminescence, which acts as a defocus term. 
Therefore, for simplicity, this focal shift is only considered for the Gaussian beam in \Cref{sec:confocal}. 

\subsection{Calculating the optical image in a wide-field microscope}\label{subsec:calc_widefield}
For the calculation of PSF in wide-field microscopy, 
the aberration function of tilted incidence [\cref{eq:Phi}] is used. 
This is due to the fact that for off-center pixels, 
the principal rays are tilted depending on the relative distance from the center.
Given that the PSF is dependent on the pixel coordinates, 
the convolution calculation is performed as follows: 
For coordinates $\vec{r} \in \mathbb{R}^2$ in the proximity of the focus, 
let us denote the light source distribution as $F(\vec{r})$. 
When the PSF is represented as a distribution $G(\tilde{r};\vec{r})$ with respect to the relative position $\tilde{r} \in \mathbb{R}^2$ and depending on the central position $\vec{r}$,
the observed intensity distribution $\tilde{F}(r)$ is represented as
\begin{equation}
\tilde{F}(\vec{r}) = \int d\vec{r'} F(\vec{r} - \vec{r'}) G(\vec{r'}; \vec{r}-\vec{r'}). \label{eq:conv}
\end{equation}

Specific results of USAF chart images compared between calculation by this model and experiments are given in \Cref{sec:widefield}.

\subsection{Execution of numerical integral}
The calculations are performed using a Julia language \cite{julia} and 
the package that performs ``h-adaptive'' multi-dimensional integration \cite{GENZ1980295,HCubature}.
This method recursively repeats the following steps until convergence:
\begin{enumerate} 
  \item subdivide the integration region, which is represented as a product of closed intervals
  \item apply the same integration rule to each region
\end{enumerate}
As this method provides intervals to evaluate the integration in an array form, 
it can be parallelized and exhibits high performance for multi-dimensional integrals\cite{berntsen1991adaptive}.

\section{Evaluation of the single NV center PSF using a confocal system}\label{sec:confocal}
We evaluate the impact of aberrations on the PL intensity distribution of NV centers. We use a confocal microscope with a moving stage to map and acquire optical images of single NV centers located at different depths from the surface.

\begin{figure}[tbp]
  \includegraphics[width = \hsize]{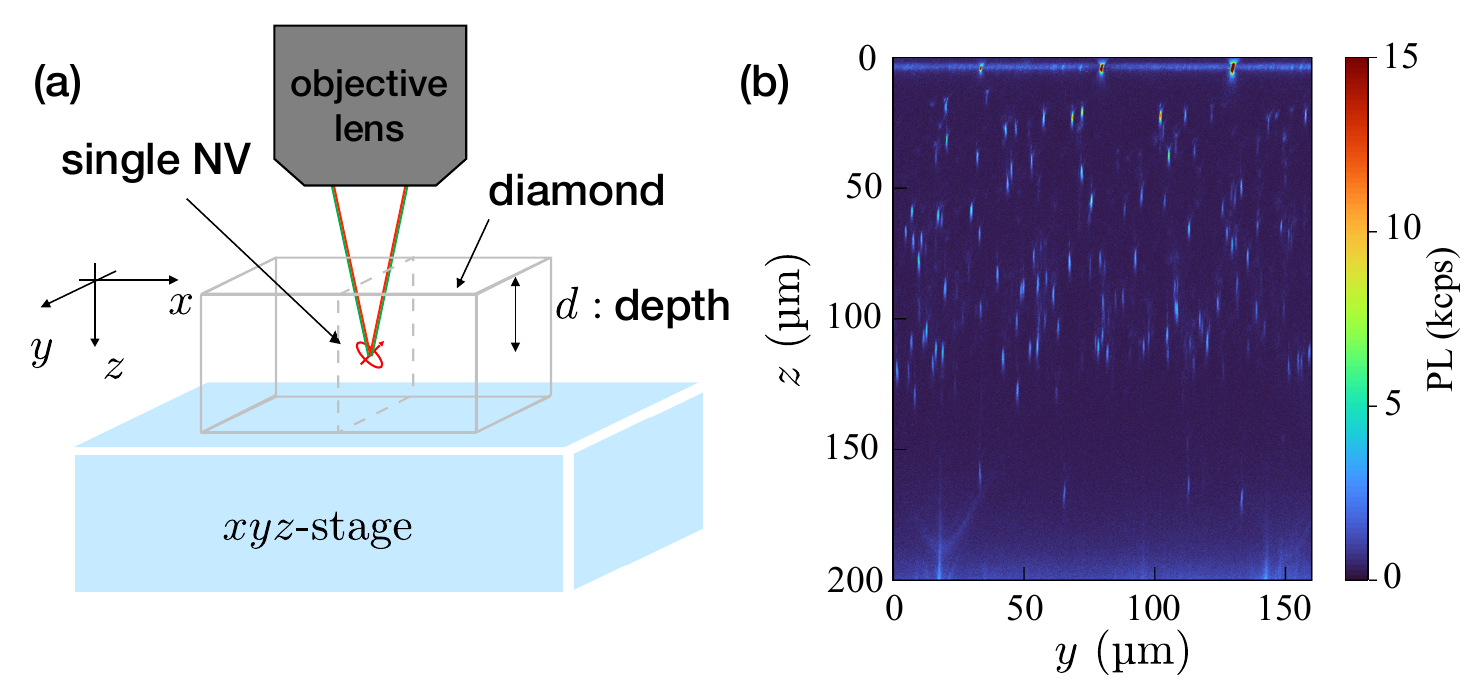}
  \caption{(a) Measurement scheme. The {$z$-axis} is taken along the optical axis toward the objective lens, and the $x$- and $y$-axes are taken perpendicular to each other in the plane normal to the {$z$-axis}.
  (b) Scanning measurement results of the cross-section in the {$yz$-plane} of a sample with single NV centers. 
  They appear as vertically elongated spots.
  The decrease in the number of NV centers in deep regions (below 50~\si{\micro\meter}) 
  is due to the difference in naturally existing quantities.}
  \label{fig:XZ_overview}
  \end{figure}
  
\begin{figure*}[tbp]
    \includegraphics[width = \hsize]{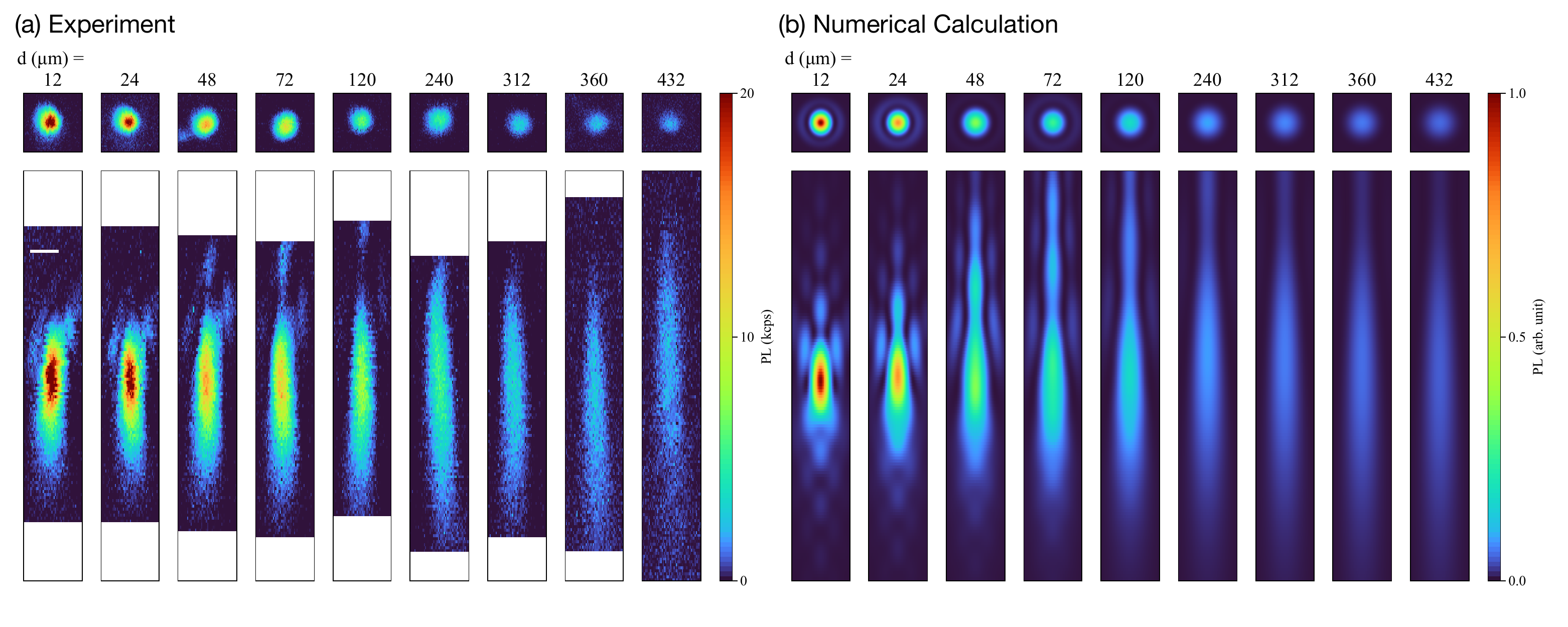}
    \caption{Comparison of measurement results of the single NV center PSF and the theoretical model. 
    The upper row represents the PL intensity distribution within the plane perpendicular to the optical axis ({$xy$-plane}), and the lower row represents the distribution within the plane parallel to the optical axis ({$xz$-plane}). 
    (a) PL intensity distribution of single NV centers located at nine different depths  $d=12~\si{\micro\meter}, 24~\si{\micro\meter}, 48~\si{\micro\meter}, 72~\si{\micro\meter}, 120~\si{\micro\meter}, 240~\si{\micro\meter}, 312~\si{\micro\meter}, 360~\si{\micro\meter}$, and $d=432~\si{\micro\meter}$.
    The white scale bar in the leftmost panel is 1~\si{\micro m}.
    (b) The calculated PSF under the same corresponding conditions. }
    \label{fig:PL_xy}
\end{figure*}

\subsection{Measurement of PL intensity distribution using confocal microscope}\label{subsec:meas_PSF}
We use single NV centers naturally present in a type \II a diamond (Element-six electronic grade, (100) substrate\footnote{Note that the PL emission from NV centers is not spherically symmetric. When calculating the diffraction integral, one might need to consider an inclination factor based on lattice symmetry.
However, we did not observe a significant inclination in the experimental PSF of the NV center. Therefore, we assume that the inclination factor can be ignored.}) as measurement targets.
The sample contains single NV centers over a wide range in depth from the surface.
A high-resolution mapping of the vicinity of some of these NV centers is 
performed by sweeping the position using a moving stage to obtain the spatial distribution of PL intensity.
We use a confocal system [\cref{fig:microscopes}(b)] built according to the design of \citet{dmisonou}.

The detailed description of our confocal system follows:
We use the objective lens of $\text{NA} = 0.8$ and a focal length of $f = 1800~\si{\micro\meter}$ on the measurement target side.
On the SPCM side, the objective lens of $\text{NA} = 0.25$ and a focal length of $f=16500~\si{\micro\meter}$ is placed, and the light is focused onto a single-mode fiber
\color{black}
(Thorlabs SM600) with a mode-field diameter of $3.6\text{--}5.3~\si{\micro\meter}$, which roughly matches $1$ Airy Unit ($1.22(\lambda = 0.7)/(\text{NA} = 0.25) \sim 3.4$). 
This design aims to balance resolution and collection eﬀiciency. 
For the detailed model numbers of the optical parts, see \citet{dmisonou}.
\color{black}
We use the $xyz$-linear stage (Sigma Tech 10~nm feedback stages FS-1020PXY for $x$- and $y$-axes, and FS-1010PZ for $z$-axis), as shown in \cref{fig:XZ_overview}(a), to map the PL intensity.
The coordinates $(x,y,z)$ in the PL intensity distribution of the subsequent data are defined by the displacement of the microscope stage. When observed using a confocal system, the diamond appears to be $1/n\sim 1/2.4$ thinner than its actual thickness due to paraxial focal shift, as previously noted in \Cref{sec:calculation}.
We estimate the thickness of the diamond transmitted, which is necessary for determining the degree of aberration, considering this focal shift.

As shown in \cref{fig:XZ_overview}(b), 
NV centers located at deeper positions appear darker and stretched in the optical axis direction due to aberration. 
To more precisely evaluate the effects of aberration, fine scans are performed for each NV center at different depths. 
The detailed scans are performed using the following steps:
\begin{enumerate}
  \item Find NV centers by mapping for fixed $z$ in wide field of view.
  \item Scan around the found NV centers location in the plane perpendicular to the optical axis.
  \item Determine the coordinates that yield the maximum intensity in {$xy$-plane}.
  \item Sweep $z$ with the $x,y$ coordinates fixed to find the coordinates that yield the maximum intensity.
  \item Perform a fine scan of each plane parallel and perpendicular to the optical axis with the obtained coordinates as the center position.
\end{enumerate}

\Cref{fig:PL_xy}(a) shows the confocal microscope image of the PL intensity distribution in the planes both perpendicular and parallel to the optical axis. We show the results for NV centers located at nine different depths between $d=12~\si{\micro\meter}$ and $d=432~\si{\micro\meter}$. From \cref{fig:PL_xy}(a), we confirm that as the NV center's position gets deeper, the peak value of the PL distribution significantly reduces. This reduction is roughly 15 times when comparing the PSF at the shallowest position ($d=12~\si{\micro\meter}$, with the least aberration) to that at the deepest ($d=432~\si{\micro\meter}$, with the most significant aberration).
Also, from \cref{fig:PL_xy}, we confirm that even if the position of the NV center gets deeper, 
the resolution in the direction perpendicular to the optical axis does not decrease much. 
In contrast, from \cref{fig:PL_xy}, we can see that as the position of the NV center gets deeper, 
the resolution in the direction parallel to the optical axis decreases. 

\Cref{fig:PL_xy}(b) shows the results of calculation using the model [\cref{eq:Ez_f}] considering the results up to \Cref{sec:calculation}. 
\color{black} Here, we assume that without the diamond, the optics are aberration-free. 
Specifically, we assume that all objectives in the system are ideal, aberration-free lenses in every aspect, including spherical aberration and chromatic aberration.
\color{black}
Both in the case of the direction parallel to the optical axis and the direction perpendicular to the optical axis, the calculated results [\cref{fig:PL_xy}(b)] reproduce the general shape of the actual PL intensity distribution [\cref{fig:PL_xy}(a)] well including the fringe, i.e., the localized peaks appearing off the central peak.

We next quantitatively evaluate the change in resolution and peak intensity due to the effects of aberration quantitatively and to compare with the calculated results.
In order to reduce noise and the effect of positioning errors, the experimental results are fitted using the following elliptical Gaussian $G(\delta x,\delta y,\delta z; I_p,\sigma,\sigma_z)$ expressed as  
\begin{equation}
G = I_p \exp(-2(\delta x^2 + \delta y^2)/~\sigma^2 - 2\delta z^2/~\sigma_z ^2).
\end{equation}
Fitting is performed after subtracting the background, which is evaluated at a sufficient distance from the PL distribution.
This fitting yields the full width at half maximum (FWHM) $\sqrt{2\log2~\sigma(\sigma_z)}$ and the PL intensity at the maximum $I_p$ as a general shape. 
For technical reasons, fitting results are first extracted from the top row of \cref{fig:PL_xy} to determine the intensity $I_p$ and the full-width at half-maximum (FWHM) $\sigma$ in the plane perpendicular to the optical axis. 
Then, using the results in the bottom row of \cref{fig:PL_xy}, fitting is performed to extract $\sigma_z$. 
To obtain the statistical mean and variances, about ten NV centers are measured at the similar depths (within 2~\si{\micro m } deviation),
and parameter estimation is performed for each.
As for the evaluation of the calculation results, 
the peak intensity is determined by finding the maximum value, 
and the FWHM is calculated from the maximum difference of the coordinates 
where the intensity is more than half of the peak intensity after removing the subpeak on the edge.

\begin{figure}[htbp]
  \includegraphics[width = \hsize]{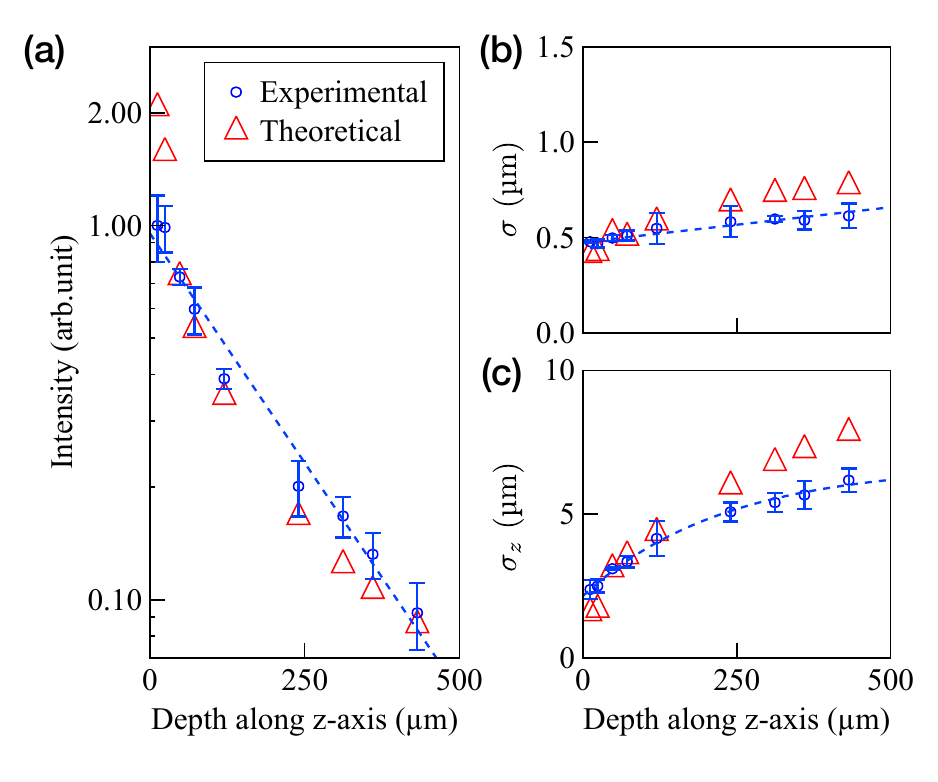}
  \caption{Evaluation of the PSF shape of each experimental result as a function of depth $d$ in the diamond. 
  (a) Peak intensity $I_p$. The experimental intensity is normalized to that for $d = 48$ \si{\micro m}. (b) Variance $\sigma$ in the {$xy$-plane}. (c) Variance $\sigma_z$ in the {$z$-axis} direction. }
  \label{fig:int_sigma}
\end{figure}

The obtained $I_p,\sigma$ and $\sigma_z$ are shown in \hyperref[fig:int_sigma]{Figs.~\ref{fig:int_sigma}(a), \ref{fig:int_sigma}(b) and \ref{fig:int_sigma}(c)}, respectively. 
For better visibility of these data, fitted curves of given forms are shown in dashed lines. 
Specifically, an exponential function is used to fit the intensity $I_p$ [\cref{fig:int_sigma}(a)] (since the data is plotted logarithmically, it appears linear), 
a linear fit is used for the FWHM $\sigma$ in the {$xy$-plane} [\cref{fig:int_sigma}(b)], and an exponential function is used for the FWHM $~\sigma_z$ in the {$xz$-plane} [\cref{fig:int_sigma}(c)].
The intensity of the numerical result is normalized based on the ratio between the experimental and calculation results for $d = 48$ \si{\micro m} (not for $d = 12$ \si{\micro m} or $d = 24$ \si{\micro m}), following subsequent discussions.

These experimental results are generally well reproduced in the calculations. 
Upon closer observation, in regions with more significant aberrations, both the FWHM in the plane parallel to the optical axis and the FWHM in the plane perpendicular to the optical axis are slightly larger in the calculated results than in the experimental results. 
Considering the high reproducibility of the numerical calculations in the experimental results of \cref{fig:PL_xy}, 
this slight difference may not be considered essential.
Instead, this result can be attributed to the fact that the Gaussian used for peak extraction does not necessarily 
represent the actual shape well, especially when the significant aberration exists.

As for the intensity $I_p$, the observed behavior for $d > 50~\si{\micro\meter}$, where the aberration is large, is consistent with the calculation. 
However, its decrease from $d=12~\si{\micro\meter}$ to $24~\si{\micro\meter}$ to $48~\si{\micro\meter}$, where the aberration is relatively weak,
is much larger in the calculation than in the experimental results.
Considering that the FWHM of the calculation is smaller than the measured results in the same region, it is reasonable to assume that the actual measurement system already contains a finite degree of aberration (possibly due to inevitable finite misalignment) and the deviation in intensity scaling can be attributed to this error.
The expression of the diffraction integral including aberrations also depends on the NA, and the larger the NA, the greater the impact of aberrations.

\subsection{Effect of aberration on excitation light intensity} \label{subsec:laser_intensity}

The effect of aberration on the excitation laser itself should also be taken into account. As noted in \Cref{subsec:calc_confocal}, the defocus term arises due to the paraxial focal shift caused by the wavelength-dependent refractive index $n(\lambda)$.
This reduction in the excitation light intensity linearly reduces the excitation rate $R$ of an NV center. 
As a result, both the PL intensity and the initialization rate, which are dependent on $R$, experience a decrease, leading to reduced sensitivity.

We experimentally measure the time-resolved PL intensity of single NV centers located at various depths. 
\footnote{This measurement of the excitation process requires pulse manipulation, so a system different from the one used in \Cref{subsec:meas_PSF} is used. This system uses a Galvano mirror to scan the measurement region that may result in a tilted incidence.
However, since the spot is measured near the sweep center of the mirror, the incident is considered to be sufficiently vertical. 
Also, note that using a different system entails that some optical parameters, especially the properties of the pinhole and the objective lens, are somewhat different from the PSF measurement system above.}
The process of optical transition is reflected in the time-resolved PL intensity of an NV center during laser irradiation. 
\begin{figure}[tbp]
  \includegraphics[width = \hsize]{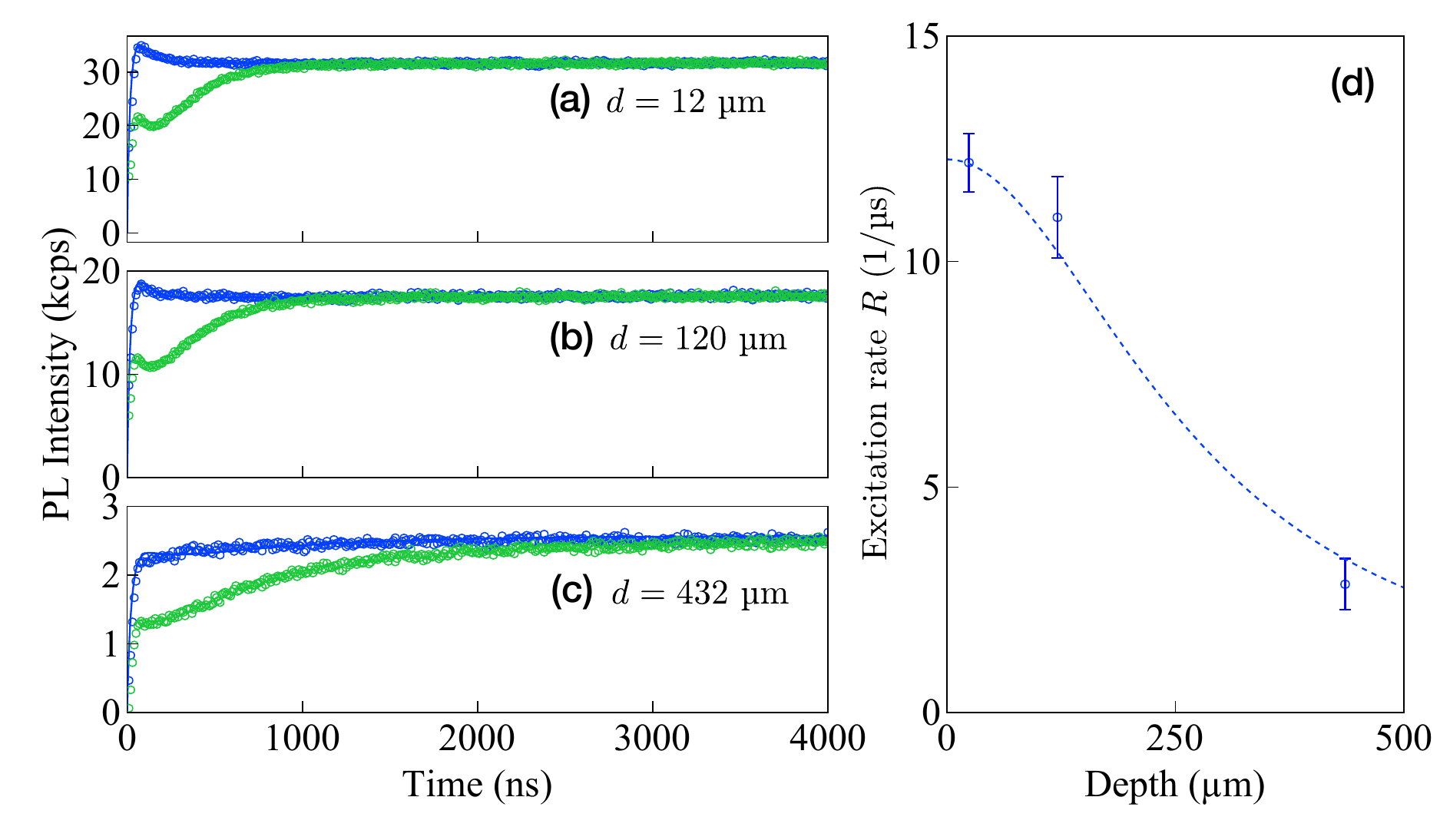}
  \caption{(a-c) Time-resolved PL intensity from the NV center. Blue lines denote the results of state $\ket{0}$, and 
  green lines denote the results of state $\ket{1}$. Results are shown at depths of (a) 12~\si{\micro m}, (b) 120~\si{\micro m}, and (c) 432~\si{\micro m}, respectively. To realize the $\ket{m_S = 1}$ state, we apply an adiabatic inversion pulse to the NV centers. 
  (d) The thickness dependence of the excitation rate estimated from (a-c) by \cref{eq:rate_sol,eq:intensity}. The blue circles represent the obtained data and the error bars are calculated by the fitting confidence interval. The dotted line is the expected curve that accounts only for the paraxial focal shift of Gaussian beam.
  }
  \label{fig:PL_fit}
\end{figure}

First, we explain the optical transition model of NV centers.
We use the minimal model representing the spin initialization to the $m_S = 0$ state\cite{steiner2010universal,robledo2011spin,doherty2013,gupta2016}, or the so-called five-level model.
Here, the states of the NV center can be simplified into five levels: two spin levels~$\ket{0}$ and $\ket{1}$, for both of the ground and excited triplet states, and one level for the singlet state. 
The process of optical transition can be modeled as a stochastic differential equation\cite{robledo2011spin}, which is complicated and hard to solve.
However, the time change of population at each level as statistical averages in time domain can be simplified into a continuous-time Markov process, and follows the linear ordinary differential equation\cite{gardiner1985}, or the so-called rate equation\cite{robledo2011spin}.

Let $ \vec{p} = [p_{g0},p_{g1}, p_{e0},p_{e1}, p_s]^\intercal$ be the population of states that exist statistically on average at each time. 
We denote the population of the ground states of $\ket{m_S = 0,1}$ as $p_{g0},p_{g1}$, 
the population of the excited states of $\ket{m_S = 0,1}$ as $p_{e0},p_{e1}$, 
and the population of the singlet state as $p_s$, respectively. 
The rate equation for time-dependent population $\vec{p}(t)$ is given by
\begin{align}
  \dv{t} \vec{p}(t) &= A\, \vec{p}(t), \\
  A &=  \begin{psmallmatrix}
      -R & 0 & \gamma & 0 & D_0\\
      0 & -R & 0 & \gamma & D_1\\
      R & 0 & -(S_0 + \gamma) & 0 & 0 \\
      0 & R & 0 & -(S_1 + \gamma) & 0 \\
      0 & 0 & S_0 & S_1 & -(D_0 + D_1)\\
  \end{psmallmatrix},
      \label{eq:rate}
\end{align}
where $A$ is the transition rate matrix, $R$ is the transition rate from the ground triplet to the excited triplet, 
$\gamma$ is the transition rate from the excited triplet to the ground triplet by direct transition, 
$S_0$, $S_1$ are the transition rates from the excited state of $\ket{m_S = 0}$ and $\ket{m_S = 1}$ to the singlet state, 
and $D_0$, $D_1$ correspond to the transition rate from the singlet state to the ground state of $\ket{m_S = 0}$ and $\ket{m_S = 1}$, respectively.
The solution can be explicitly expressed using a matrix exponential function as 
\begin{equation}
  \vec{p}(t) = e^{At} \vec{p}(t=0). \label{eq:rate_sol}
\end{equation}
The experimentally observable PL intensity, which is proportional to the total relaxation rate directly down to the ground triplet from the excited triplet, can be expressed as 
\begin{equation}
  I(t) = C\gamma (p_{e0}(t) + p_{e1}(t)), \label{eq:intensity}
\end{equation}
where the coefficient $C$ is the collection efficiency, i.e., the parameter introduced to consider that not all PL emissions by direct transition can be observed in the measurement due to the constraints of the optical system such as the objective lens, and only a part of them enters the detector.

The initial states are prepared as\cite{gupta2016},
\begin{equation}
  \vec{p_0}(t=0) = \mqty(1-2\zeta\\ 2\zeta\\ 0\\0\\0),\quad \vec{p_1}(t=0) = \mqty(\zeta\\ 1-\zeta\\ 0\\0\\0).
  \label{eq:polarization}
\end{equation}
Here, $ \zeta$ is the initial polarization ratio, 
which represents the degree to which the ground state is finally polarized to $ \ket{m_S = 0}$ by the sufficient time of excitation light irradiation.
$\vec{p}_{0(1)}(t)$ is the time-dependent population starting from each initial state $\ket{m_S = 0(1)}$.
Fitted by this model, each parameter of $A$ in \cref{eq:rate} can be estimated from the experimental results.
\footnote{
Since there are seven parameters for the two observable curves of the time-resolved PL intensity, and $C$ and $\gamma$ are represented by direct products and have covariance, parameter estimation by least squares results in large covariance. 
In this case, if all parameters are changed and optimization is performed, the confidence interval of the parameters becomes large.
Therefore, the optimization is performed assuming that only the collection efficiency $C$, the initial polarization ratio $ \zeta$, and the excitation rate $R$ of the estimated parameters differ for each NV center at different depths. 
The other parameters are fixed to common values with each NV center.
}

\hyperref[fig:PL_fit]{Figures \ref{fig:PL_fit}(a), \ref{fig:PL_fit}(b), and \ref{fig:PL_fit}(c)} shows the experimental result of time-resolved PL intensity of single NV centers at different depth $d$.
Each curve is obtained by summing up at each time bin of $10$ ns.
As can be seen from \hyperref[fig:PL_fit]{Figs. \ref{fig:PL_fit}(a), \ref{fig:PL_fit}(b), and \ref{fig:PL_fit}(c)}, the time required for initialization, i.e., the time until the deviation in the PL intensity of the two states $\ket{m_S = 0, 1}$ disappears, changes with depth. 
This reflects the fact that the excitation light intensity irradiated to the NV center decreases with depth due to aberration. 
This phenomenon is characterized by the excitation rate $R$, which is a parameter proportional to the excitation light intensity. 
\Cref{fig:PL_fit}(d) shows the dependence of $R$ on the depth, which is obtained by extracting parameters by the method mentioned earlier in \Cref{subsec:laser_intensity}. 
The excitation rate $R$ does not decrease much for the thickness about $120~\si{\micro\meter}$, 
but drops dramatically at $432~\si{\micro\meter}$. 

This decrease in excitation light is attributed to the focal shift.
The shift is caused by the change in the refractive index of the diamond between the wavelength of the excitation light, 
532~nm, and the peak wavelength of the NV center's luminescence spectrum, around $700$~nm.
The refractive index of a diamond\cite{turri2017index}, depending on the transmitting light's wavelength, is given as 
\footnote{
  Although \citet{turri2017index} does not square the value $112.5$, it contradicts the measurement results given in this paper if $112.5$ is not squared, 
  thus we use a complementary function modified to better explain the measurement results of the paper.} 
\begin{equation}
n(\lambda) = \sqrt{1 + \frac{4.658\lambda^2 }{\lambda^2 - 112.5^2}}.\label{eq:refractive}
\end{equation}
Since \cref{eq:refractive} changes dramatically around visible light wavelength, the defocus effect due to this change is large. 
Specifically, the focus of light with a wavelength corresponding to the excitation light, $\lambda = 532~\si{\nano\meter}$, 
appears to shift up by,
\begin{equation}
  0.00336\times d~(\si{\micro\meter}), \label{eq:defocus}
\end{equation}
more than those for the wavelength of the NV center's PL emission $\lambda\sim 700~\si{\nano\meter}$. 
Considering this, we assume that the laser excitation light is an ideal Gaussian beam with a $1/e^2$ beam waist radius $w_0 = 0.3~\si{\micro\meter}$ (diffraction limit scale when NA is 0.8), and assume its distribution in the form of \cref{eq:beam}.
The estimation of decrease in excitation light intensity is expected assuming that the beam waist of the excitation light distribution [\cref{eq:beam}] is shifted from the focal point of the PSF of a single NV center according to \cref{eq:defocus} while the intensity distribution is kept unchanged. The estimated curve is represented by the dotted line in \cref{fig:PL_fit}(d).
This curve reproduces the experimental results well. 
In more detail, the Gaussian beam is also affected by higher-order aberration, and the beam shape itself can deviate from \cref{eq:beam}.
This can also be calculated from diffraction theory\cite{dickson1970characteristics, miks2003propagation}.

\subsection{Summary of results for confocal system}
So far, we have discussed the experimentally observed single NV center PSF in a confocal system. 
The calculation results based on the diffraction theory reproduce the measurement results well.
Specifically, the observation results are that the intensity and axial resolution decrease each time they pass through the diamond, 
while the resolution in the direction perpendicular to the optical axis does not change much.
This reflects the characteristics of spherical aberration, which describes the third order symmetric component of the aberration given in \citet{born_wolf}.
In more detail, the experimental PSF and the calculation results slightly deviate mainly in terms of intensity in the region close to the diamond surface where aberrations are relatively minor.
We attribute this minor deviation to the errors of the optical system without placing the diamond.
Based on confirmation that our model reproduces experimental results well,
in the following section, we discuss the impact of aberrations in wide-field microscopy using the same diffraction theory model.

\section{Resolution of wide-field microscopy with an image sensor}\label{sec:widefield}
We evaluate the effects of aberrations in the case of wide-field microscopy with an image sensor [\cref{fig:microscopes}(a)]. 
In this case, the incidence is generally oblique on parallel plates as shown in \cref{fig:normal_incidence}(b). 
Therefore, the resolution depends on the angle of incidence at the viewpoint, that is, the distance from the view center. 
We compare in detail the experimentally obtained wide-field image of USAF chart [\cref{fig:imaging}] and our model [\cref{eq:Ez_f}] in the case of tilted incidence.
The optical system for wide-field imaging treated in this section has $\text{NA} = 0.7$ and $f = 2000~\si{\micro m}$. 
\color{black}(Mitsutoyo MPlan Apo 100x) with a $100$ mm tube lens (Thorlabs AC254-100-B), resulting in a magnification of $50\times$. 
These lenses are assumed aberration-free, which is supported by the observation that there is no significant degradation without a diamond in resolution at the edge in Figs. 2 and 11. 
\color{black}
These parameters are fixed throughout this section.
In our case, the image sensor size should also be taken into account when considering aberration. In our wide-field microscopy optical system, we use a CMOS camera with a sensor size of $7.07~\si{mm} \times 5.3~\si{mm}$. 
\textcolor{black}{With $50\times$ magnification}, the field of view measures $141.3~\si{\micro m} \times 106.0~\si{\micro m}$.
\begin{figure}[tbp]
  \includegraphics[width = \hsize]{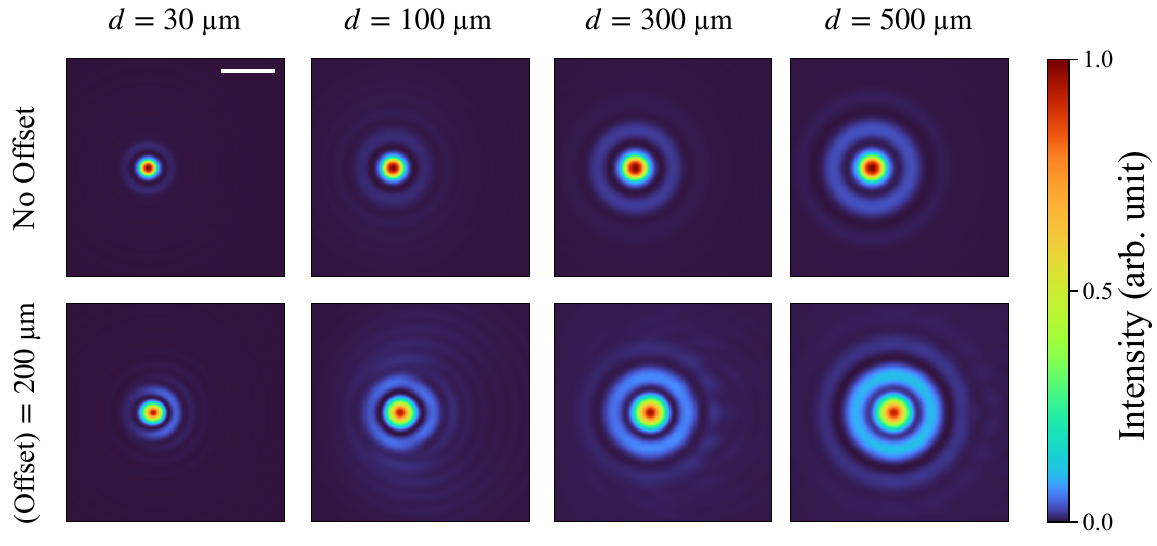}
  \caption{
    Calculated results of PSF in wide-field microscopy. The upper and lower rows show the result calculated at the view center (``No Offset'') and at the edge of the field of view (``(Offset) = 200~\si{\micro m}''), respectively.
    The intensity is normalized by the maximum intensity at each thickness when located at the center (``No Offset''), and the focus is taken at the maximum point of intensity at the view center.
  }
  \label{fig:psf_offset}
\end{figure}

\begin{figure*}[t]
  \includegraphics[width = \hsize]{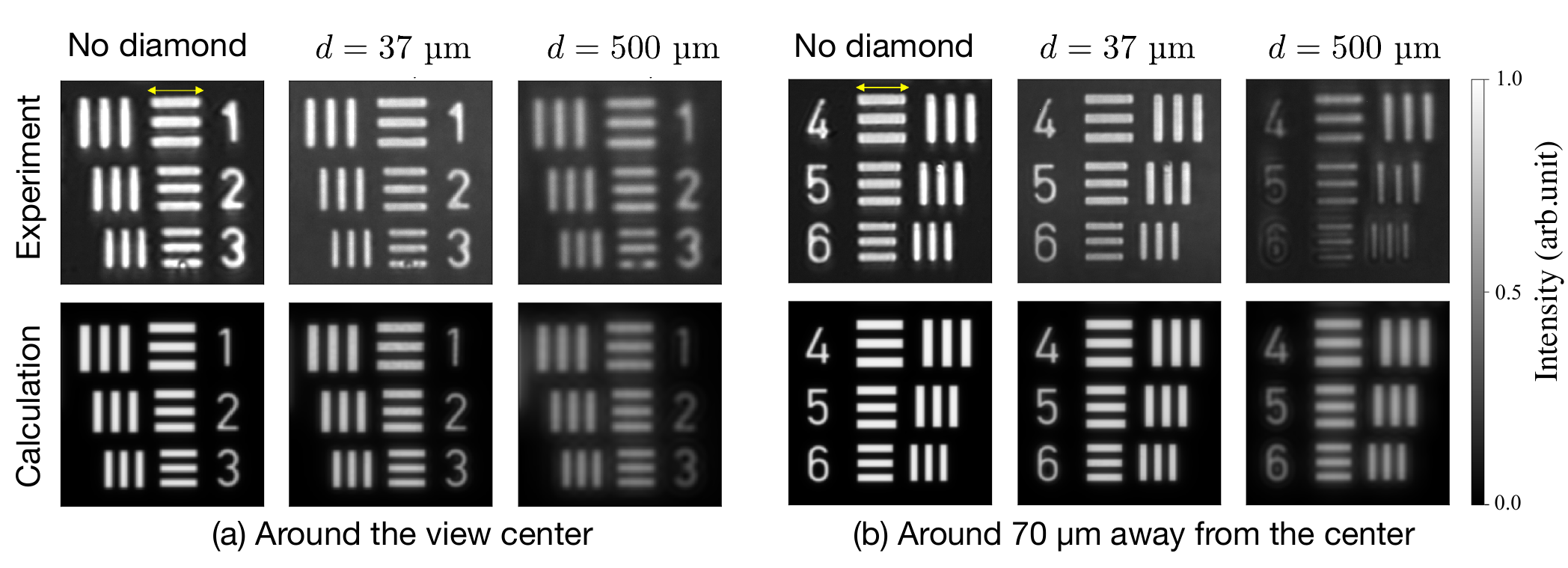}
  \caption{
    Enlarged views of the USAF chart image shown in \cref{fig:imaging} and simulation results calculated using convolution \cref{eq:conv} under corresponding conditions.
  (a) Pattern image of Group 9, Element Number 1-3 near the view center. The yellow arrow in the left-top panel indicates $4.9~\si{\micro\meter}~( = 2.5~\si{\micro m } \times 2^{-(9 + 0/6)})$.
  (b) Pattern image of Group 8, Element Number 4-6, around 70~\si{\micro m} away from the center. The yellow arrow in the left-top panel is $6.9~\si{\micro\meter}~( = 2.5~\si{\micro m }  \times 2^{-(8 + 3/6)})$.}
  \color{black}
  Here, the incident light is assumed to have randomly oriented linear polarization.
  Furthermore, there may exist a finite gap between the diamond plate surface and the USAF test chart. 
This does not affect the results, as the aberration function does not depend on this gap distance.
\cite{braat1997analytical}.
\color{black}
  \label{fig:USAF_sim}
\end{figure*}

\subsection{Optical image calculated from PSF}
\Cref{fig:psf_offset} shows the PSF calculated at the center of the field of view and a location 200~\si{\micro m} off-center for $d=30$~\si{\micro m}, 100~\si{\micro m}, 300~\si{\micro m}, and 500~\si{\micro m}.
For wide-field microscopes, the angle of the incidence at the measurement point essentially affects aberration. 
This incident angle is determined using coordinates relative to the center, taking magnification into account, and is reflected in the shape of the PSF.
The PSFs calculated by considering the above are shown in \cref{fig:psf_offset}.
The spread of the PSF increases, and the peak intensity decreases when placed far from the view center. 
Furthermore, we confirm that the intensity of the fringe increases away from the view center. 
This fringe appears asymmetrically, i.e., slightly brighter on the far side (right side in each figure), 
suggesting the contribution of coma aberration and astigmatism\cite{born_wolf}.
From \cref{fig:psf_offset}, we also confirm that even in the case of wide-field microscopy with asymmetric components of aberration due to the tilted incidence, 
the symmetrical components of the aberration function remain dominant.

Next, the USAF chart image is simulated by convolution of the obtained PSF. 
The optical image is calculated using \cref{eq:conv} with the PSF numerically obtained using the theory used in the calculation of \cref{fig:psf_offset}.
The distribution of the light source in \cref{eq:conv} is obtained from vectorizing \cref{fig:imaging}(a) with edge detection. 
We then generate raster data rescaled to the pixel size of the calculated PSF.

\Cref{fig:USAF_sim}(a) shows the calculation results of the optical image at the corresponding position 
and the measurement results near the center of the images shown in \hyperref[fig:imaging]{Figs. \ref{fig:imaging}(a), \ref{fig:imaging}(b), and \ref{fig:imaging}(c)}. 
In contrast, \cref{fig:USAF_sim}(b) shows the calculated and experimental results near the edge of the field of view (near the element number 4-6 of the USAF chart) of \hyperref[fig:imaging]{Figs. \ref{fig:imaging}(a), \ref{fig:imaging}(b), and \ref{fig:imaging}(c)}.
The experimental results have been processed to exclude a certain background emission to eliminate diffusive reflections from the diamond.

We first compare near the view center [\cref{fig:USAF_sim}(a)]. 
When a thin diamond (37~\si{\micro m}) is placed, it can be seen that the optical resolution is almost unchanged compared to the case when no diamond is placed. 
On the other hand, when a 500~\si{\micro m} diamond (which is commercially available and typical) is placed, the resolution is reduced as mentioned in \Cref{sec:intro}. 
The numerical calculation appropriately reproduces this behavior.

Next, we compare at the edge of the field of view (around 70~\si{\micro m} away from the view center) [\cref{fig:USAF_sim}(b)].
It is confirmed that even if a thin (37~\si{\micro m}) diamond is placed, 
the resolution does not decrease much, and the numerical calculation also reproduces this. 
In contrast, when a 500 $\si{\micro m}$ diamond is placed, 
the actual optical image has a lower resolution than the case at the view center, and a bleeding effect appears to surround the pattern.
This optical image suggests the presence of the large fringe that appears especially when there is an offset from the center in \cref{fig:psf_offset}. 
From the above, we confirm that the diffraction integral also reproduces the change in the optical image in the case of a wide-field microscope.

We also find that the contrast of the images decreases when diamonds are placed. The reduced contrast can partly be attributed to the aberration, as suggested by \cref{fig:psf_offset}. However, other factors could reduce contrast as well. The actual optical image results from the reflection of the finite laser spot. We consider this reflection as if the PL emission is coming from a light source with the same wavelength of $532$~nm, which exists in close contact with the back surface of the diamond. When a 500~\si{\micro m} diamond is placed on the USAF chart, the laser spot size is modified due to the effect of refraction, which might lead to a decrease in reflection intensity. We also account for the background intensity, which increases when a diamond is placed due to diffusive reflection. Given these factors, we cannot quantitatively discuss the contrast in light intensity between the signal and the background.

\subsection{Achievement of diffraction limit}

We discuss the thickness of the diamond allowed to achieve the diffraction limit, using the Strehl ratio at the focus defined in \Cref{subsec:tolerance}. 
We employ a linear search to find the thickness $d$ that satisfies the condition $S>0.8$ for $\lambda = 700~\si{\nano\meter}$.
We provide a list of diamond thicknesses that achieve the diffraction limit, corresponding to various distances from the view center as follows:
\begin{equation}  \label{eq:limit}
  d = \begin{cases}
    \textcolor{black}{41}~\si{\micro\meter} & \text{when offset} = 0~\si{\micro\meter}\\
    \textcolor{black}{31}~\si{\micro\meter} & \text{when offset} = 100~\si{\micro\meter}\\
    \textcolor{black}{20}~\si{\micro\meter} & \text{when offset} = 200~\si{\micro\meter}
  \end{cases}
\end{equation}
We can briefly justify our evaluation criterion based on \cref{fig:USAF_sim}. 
From the list [\cref{eq:limit}], the optical image transmitted through a {37~\si{\micro m} thick} diamond (thinner than \textcolor{black}{$41$} \si{\micro m}) should achieve the diffraction limit within a range close to the view center. 
This conclusion is consistent with the fact that there is little degradation in the optical image [see Figs.~\hyperref[fig:imaging]{\ref{fig:imaging}(b)} and \hyperref[fig:USAF_sim]{\ref{fig:USAF_sim}}], compared to the case where no diamond is present [see Figs. \hyperref[fig:imaging]{\ref{fig:imaging}(a)} and \hyperref[fig:USAF_sim]{\ref{fig:USAF_sim}}].

We discuss the resolution theoretically achieved under the diffraction limit.
The PL emission of the ensemble NV center can be considered as incoherent. 
Therefore, we use Rayleigh's criterion\cite{born_wolf}, 
which is widely used for evaluating the resolution in the case of incoherent light sources. 
In Rayleigh's criterion, the resolution is given as $0.61 \frac{\lambda}{\text{NA}}$.
Thus, for the light of wavelength $\lambda \simeq 650\text{--}750~\si{\nano\meter}$, 
which is the range of the PL spectrum of the NV center,
the resolution is,
\begin{equation}
  566~\si{\nano\meter}\lesssim 0.61 \frac{\lambda}{\text{NA}} \lesssim 654~\si{\nano\meter}.
\end{equation}

Therefore, it is concluded that to achieve the diffraction limit, indicating around $500\text{--}600$~nm resolution, it is necessary to keep the diamond thickness down to about $30~\si{\micro\meter}$. 
It is possible to manufacture diamonds of such thickness. There are also precedent studies for sensing applications using such thin diamonds
\cite{schlussel2018}
\footnote{Note that in this study\cite{schlussel2018}, thinning the diamond is mainly for getting closer to the measurement target, and it might not be aimed at reducing aberrations.}.
However, diamonds have the highest Mohs hardness of 10, and thinning them down to tens of $\si{\micro m}$ \cite{appel2016fabrication} is technically difficult. 
In addition, in measurements using ensemble NV centers, since the target and the diamond are often directly contacted,  it is also convenient if the diamond has a certain thickness in some cases (for example, to establish thermal contact in cryogenic application\cite{scott2020laser,sn2023}).
Therefore, in practice, it is useful to relax the conditions and determine the thickness of the diamond 
according to the resolution required for the measurement target. 
In the next subsection, the expected resolution for each thickness is presented.

\subsection{Evaluation of optical resolution for given thickness of diamond}
We evaluate the achievable optical resolution for a given diamond thickness, taking aberrations into account.
As mentioned in \Cref{subsec:planeparallel}, the degree of aberration is proportional to the diamond's thickness.
Therefore, the achieved resolution increases monotonically with respect to the thickness, and 
we can inversely determine the allowable thickness of the diamond to meet the desired optical resolution. 
The Rayleigh criterion corresponds to the full width at 40\% of the peak intensity of the Airy Disk\cite{born_wolf}, which is ideally shaped as a sinc function. However, in the general case of vectorial diffraction integral, the PSF shapes deviate from the sinc function, as illustrated in \cref{fig:PL_xy}.
Thus, we employ the generalized Rayleigh criterion defined as 
the diameter $R$ of the disk region $D$ that encompasses the region where the intensity $I(\vec{r})$ at a point $\vec{r}$ satisfies
\begin{equation}
  I(\vec{r})/I_\mathrm{max} > 0.4 
\end{equation}
with respect to the maximum intensity $I_\text{max}$, where $I_\text{max}$ is the maximum emission intensity. 

\begin{figure}[tbp]
  \includegraphics[width = \hsize]{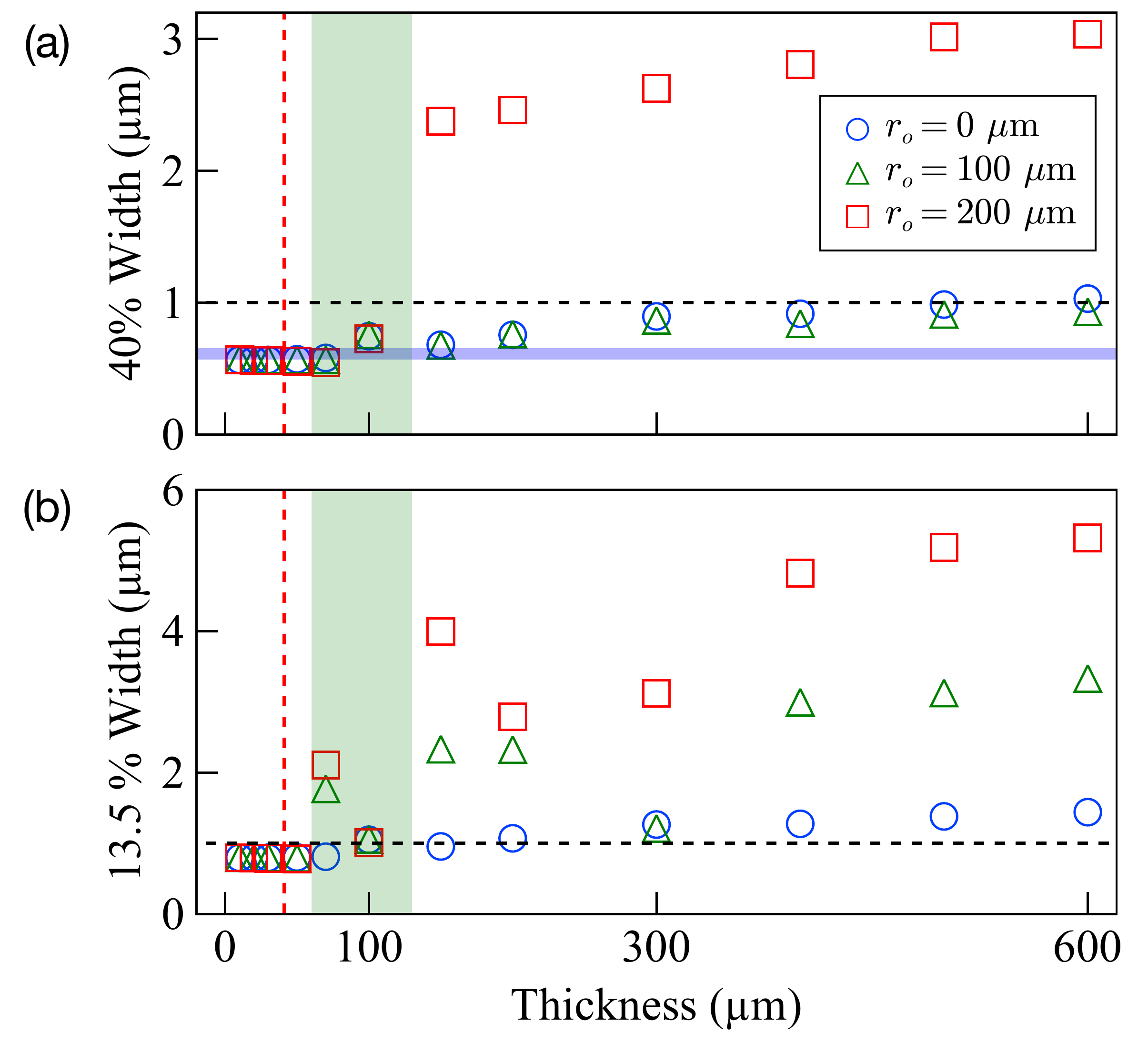}
  \caption{
    Diamond thickness dependence on the expected resolution achievable 
    \color{black}
    throughout for three specific different view ranges, i.e., offsets $r_o = 0, 100, 200~\si{\micro\meter}$ from the center of the field of view.
    \color{black}
    (a) A 40\% width, similar to the Rayleigh criterion.  (b) $13.5\% \sim 1/e^2$ value width.  Furthermore, the threshold for $1 \si{\micro m}$ resolution for each criterion is indicated
    by a black dashed line, while the thickness achieving the diffraction limit is marked by a red dashed line.}
  \label{fig:resolution}
\end{figure}

The resolution at each view offset $r_o$ calculated using this criterion is shown in \cref{fig:resolution}(a).

This result indicates that with a diamond not exceeding $500~\si{\micro m}$ in thickness, 
a resolution of $1~\si{\micro m}$ can be achieved within a field of view of $100~\si{\micro m}$ (comparable to \cref{fig:imaging}).
On the other hand, we find that the resolution decreases drastically to about $3~\si{\micro m}$ when the field of view is approximately $200~\si{\micro m}$, indicating that the intensity of the fringes seen in \cref{fig:psf_offset} becomes large to cover the $40\%$ value threshold in the area with significant aberrations, as one can also see in \cref{fig:imaging}(c).

Therefore, when the fringe takes a large value, the evaluation may be strongly affected by the threshold. 
In order to confirm the versatility of the criterion, we next re-evaluate the resolution 
by imposing more stringent conditions. 
Another criterion frequently used in optics is the $1/e^2$ width \cite{born_wolf}. 
This corresponds to a threshold value of about $13.5\%$. 
The resolution evaluated by putting the threshold to $1/e^2 \sim 13.5\%$ is shown in \cref{fig:resolution}(b). 
In this case, the resolution drops to $2~\si{\micro m}$ around the green area of $70 - 100~\si{\micro m}$.

From these results, in order to stably achieve a resolution of 1~\si{\micro m} throughout the field of view of a wide-field microscope, 
taking into account the blurring due to the fringe, it is necessary to make the diamond thinner than at least $100~\si{\micro m}$. 
Strictly speaking, if the thickness is not less than $70~\si{\micro m}$, 
the resolution is worse than 1~\si{\micro m} at 100~\si{\micro m} away from the view center.
However, at this time, the peak size of the fringe is 22\% of maximum intensity, which exceeds the threshold merely by 7\%. 
Therefore, in practice, there is little concern if the diamond thickness ranges between $70\text{--}100~\si{\micro m}$, and the resolution can be practically considered to be 1~\si{\micro m}. 
Additionally, if one only observes around the view center, the tilted incidence is almost negligible and the resolution stays around 1.5~\si{\micro m} even if one uses a commercially available 500~\si{\micro m} diamond. Note that the resolution under aberration strongly depends on the design of the optical system, especially on the NA, focal depth, magnification, and image sensor size.

\subsection{Summary of the results for wide-field microscopy}
We summarize the achievable resolution for a given thickness of diamond sample.
By applying the resolution evaluation using diffraction integration to wide-field microscopy, the diamond thickness necessary to achieve the diffraction limit at the view center is about 41~\si{\micro m} based on the Strehl ratio criterion. 
Furthermore, using the generalized Rayleigh criterion (40\% value diameter), we find that the diamond thickness to achieve a resolution of $1~\si{\micro m}$ across the entire field of view (around 100~\si{\micro m} radius) is $70 - 100~\si{\micro m}$. 
If the thickness exceeds this value, the resolution deteriorates to worse than approximately 2~\si{\micro m}, primarily at the edge of the field of view, due to the effect of fringes. When using a commercially available diamond with a typical thickness of 500~\si{\micro m}, 
although the resolution is degraded to around 3~\si{\micro m} for the whole field of view, 
we confirm that it stays at around 1.5~\si{\micro m} around the view center.

\section{Conclusion}\label{sec:conclusion}
To conclude, we quantitatively investigate the impact of aberrations on the QDM technique. We first introduce a rigorous expression of the diffraction integral, incorporating aberrations, for both confocal and wide-field microscopes. 
Next, we perform experimental verification of the PSF by mapping single NV centers through diamonds of various thicknesses using a confocal microscope. 
The results indicate that aberrations primarily affect longitudinal resolution and reduce peak intensity, aligning well with the diffraction integral model. 
This evaluation is also extended to wide-field microscopy, investigating the maximum diamond thickness required to achieve specific resolutions. 
We find that a diamond thickness of less than 40 \si{\micro m} is necessary to reach the diffraction limit, and less than 100 \si{\micro m} is required for achieving a resolution of 1 \si{\micro m} across the entire field of view. 
A commercial diamond with a thickness of 500 \si{\micro m} offers approximately 1.5 \si{\micro m} resolution near the center and 3 \si{\micro m} at the edge.

Our work has introduced a rigorous method for quantitatively gauging resolution in optical systems impacted by aberrations resulting from ray transmission through diamonds. This offers a road-map for creating QDM optics with enhanced resolution and sensitivity. 
Furthermore, we confirm that the expressions of the diffraction integral used in this study\cite{wolf1959electromagnetic, braat2003extended, van2010extended} 
well reproduce the experimental results of the PSF in the optical system with a diamond. 
This suggests that the predetermined PSF obtained by this model will apply to the localization techniques\cite{betzig2006imaging, hess2006ultra-high, rust2006sub} and image refinement techniques using deconvolution.

We thank K. M. Itoh (Keio University) for providing the confocal microscope system.
This work is partially supported by JST, CREST Grant Number JPMJCR23I2, Japan; JSPS Grants-in-Aid for Scientific Research (Nos.~JP23H01103, JP22K03524, JP22KJ1059, JP22J21412, and JP19H05826); Next Generation Artificial Intelligence Research Center at the University of Tokyo; ``Advanced Research Infrastructure for Materials and Nanotechnology in Japan (ARIM)'' of the Ministry of Education, Culture, Sports, Science and Technorogy (MEXT), Proposal Number JPMXP1222UT1131; Daikin Industries, Ltd; FoPM, WINGS Program, the University of Tokyo; Kondo Memorial Foundation; the Cooperative Research Project of RIEC, Tohoku University. S.N. is supported by the Forefront Physics and Mathematics Program to Drive Transformation (FoPM), WINGS Program, and JSR Fellowship, the University of Tokyo.
% \bibliography{bibmain.bib}
%aipnum4-2.bst 2019-01-14 (MD) hand-edited version of apsrev4-1.bst

\end{document}